\documentclass[aps,pra,twocolumn,superscriptaddress,longbibliography]{revtex4-2}

\usepackage{lipsum}
\usepackage{graphicx}
\usepackage{amsmath}
\usepackage{amssymb}
\usepackage{color}
\usepackage[normalem]{ulem}
\usepackage[utf8]{inputenc}
\usepackage[colorlinks = true,
            allcolors = {Blue}]{hyperref}

\usepackage{environ}
\usepackage{comment}
\usepackage[dvipsnames]{xcolor}

\newcommand{\Dt}{\ensuremath{\delta t}}
\newcommand{\op}[1]{\hat{#1}}
\newcommand{\abs}[1]{\ensuremath{|#1|}}
\newcommand{\norm}[1]{\left\lVert#1\right\rVert}

\newcommand{\expect}[1]{\langle #1 \rangle}

\newcommand{\e}[1]{\,e^{#1}}
\newcommand{\E}[1]{\exp\left( #1 \right)}
\newcommand{\LL}{\mathcal{L}}
\newcommand{\pare}[1]{\left( {#1} \right)}
\newcommand{\spare}[1]{\left[ {#1} \right]}

\newcommand{\id}{1\!\!1}

\begin{document}
\title{Tensor network simulation of chains of non-Markovian open quantum systems}
\author{Gerald E. Fux}
\affiliation{SUPA, School of Physics and Astronomy, University of St Andrews, St Andrews, KY16 9SS, United Kingdom}
\affiliation{Abdus Salam International Center for Theoretical Physics (ICTP), Strada Costiera 11, 34151 Trieste, Italy}
\author{Dainius Kilda}
\affiliation{Max-Planck-Institut f\"ur Quantenoptik, Hans-Kopfermann-Str. 1, D-85748 Garching, Germany}
\author{Brendon W. Lovett}
\affiliation{SUPA, School of Physics and Astronomy, University of St Andrews, St Andrews, KY16 9SS, United Kingdom}
\author{Jonathan Keeling}
\affiliation{SUPA, School of Physics and Astronomy, University of St Andrews, St Andrews, KY16 9SS, United Kingdom}

\date{\today}

\begin{abstract}
We introduce a general numerical method to compute dynamics and multi-time correlations of chains of quantum systems, where each system may couple strongly to a structured environment.
The method combines the process tensor formalism for general (possibly non-Markovian) open quantum systems with time evolving block decimation (TEBD) for 1D chains.
It systematically reduces the numerical complexity originating from system-environment correlations before integrating them into the full many-body problem, making a wide range of applications numerically feasible.
We illustrate the power of this method by studying two examples.  First, we study the thermalization of individual spins of a short XYZ Heisenberg chain with strongly coupled thermal leads.
Our results confirm the complete thermalization of the chain when coupled to a single bath, and reveal distinct effective temperatures in low, mid, and high frequency regimes when the chain is placed between a hot and a cold bath.
Second, we study the dynamics of diffusion in an longer XY chain, when each site couples to its own bath.
\end{abstract}

\maketitle


\section{Introduction}
A key challenge in the field of open quantum systems is the description of systems that couple strongly to structured environments.
In general, these systems do not admit time-local (Markovian) equations of motion and thus make a non-Markovian description necessary~\cite{Breuer2002, DeVega2017}.
Such non-Markovian open quantum systems generally suffer from exponential growth of complexity with the memory time of the environment in a very similar way as the complexity of many-body quantum systems grows with the number of relevant sites.
There is even a range of interesting physical scenarios that include both many-body quantum systems \emph{and} strongly coupled structured environments~\cite{Blanter2000, Agrait2003, Giazotto2006, Losego2012, Widawsky2013}. Such scenarios are of importance for fundamental research, such as the study of strong coupling quantum thermodynamics~\cite{Nicolin2011, Horodecki2013, Vinjanampathy2016, Seifert2016, Binder2019, Talkner2020}, as well as technological and biological applications~\cite{Bose2003, Wojcik2005, Engel2007, Lambert2013, Motlagh2014, Mitchison2019, Cao2020}.
However, almost all methods for the study of many-body systems only consider closed or Markovian dynamics, while methods for the study of non-Markovian open quantum systems are generally restricted to small system sizes (we briefly review the exceptions below~\cite{Prior2010, Chin2010, Makri2018a, Makri2018, Suzuki2019, Mascherpa2019, Purkayastha2020, Makri2021, Kundu2021, Kundu2021a, Flannigan2021, Bose2021}).

Notably, the most commonly applied approaches for open quantum systems, such as the time-convolutionless, Nakajima-Zwanzig, and Gorini-Kossakowski-Sudarshan-Lindblad (GKSL) master equations~\cite{Breuer2002}, aim at correctly describing the reduced system dynamics, but are in general unsuitable for obtaining correct multi-time correlations.
In many experiments the observed quantities are, however, related to multi-time correlations, such as the fluorescence and absorption spectra in molecular spectroscopy, and bunching and anti-bunching of photons in quantum optics experiments.
Even the exact knowledge of the evolution of the reduced density matrix is not sufficient to correctly describe multi-time correlations.
The common approach to invoke the so called ``quantum regression theorem''~\cite{Lax1963} relies on the Born approximation and thus assumes weak coupling between the system and environment~\cite{Ford1996, Guarnieri2014}.
Particularly interesting scenarios, however, involve strong coupling to both the neighboring parts of a many-body system and the continuous set of modes of a thermal bath.
In such cases, correlations between the system and environment play an important role for the correct computation of the two-time correlations but cannot be encoded in the reduced density matrix alone.

In this paper we introduce a numerical method that enables the computation of the dynamics and multi-time correlations of chains of non-Markovian open quantum systems.
As discussed below, it is based on a representation of the process tensor in a matrix product operator form \mbox{(PT-MPO)} which encodes the complex system-environment correlations and allows the compression of the influence of the environment to capture the most physically relevant sector of the exponentially large state space~\cite{Jorgensen2019, Strathearn2019, Lerose2020, Fux2021, Cygorek2021,Sonner2021, Ye2021, White2021, Thoenniss2022,  Thoenniss2022efficient,Ng2022real,OQuPy2022}.
We illustrate the use of this method by studying two examples.
We first consider 
two-time correlations of an XYZ spin chain strongly coupled to its environment in both equilibrium and non-equilibrium scenarios.
We verify that for coupling to a single bath the computed two-time correlations obey the fluctuation-dissipation theorem (FDT) and contrast this to the failure of a widely-used approximate approach.
For two baths at different temperatures, we can define a frequency-dependent effective temperature $T(\omega)$ and identify different behavior in low, mid, and high frequency regimes.
In addition to this, we study a 21-site anisotropic XY spin chain with a non-Markovian bath attached to every site each.
In particular, we compute the dynamics of an initial excitation in the middle of the chain and find diffusive behavior with diffusion rates depending on the XY-anisotropy and the coupling strength to the baths.

The remainder of the paper is organized as follows.
Section~\ref{sec:pt-tebd} provides an overview of the PT-MPO approach and its extension to one-dimensional chains.
Section~\ref{sec:xyz} discusses the application of this method to the XYZ model with leads,  while section~\ref{sec:xy} discusses the application to diffusion in the XY model.
Appendices provide further details on the algorithm, and on the details of computations for the two examples.


\section{Process tensors and TEBD}
\label{sec:pt-tebd}

\subsection{Overview of method}
In this section we outline
a general numerical tensor network method to study multi-time correlations of chains of open quantum systems---further details of the algorithm and its performance are given in appendix~\ref{app:PT-TEBD}.
Each system may individually couple strongly to a structured environment.
The method is applicable to a wide variety of different environments such as boson, fermion, and spin baths.
The only restriction on the environments is that the associated PT-MPO can be efficiently constructed.

The PT-MPO is based on the process tensor formalism which is a general operational approach to non-Markovian open quantum systems.
Its central object---the process tensor (PT)~\cite{Pollock2018}---is a multi linear map from the set of all possible system control operation sequences to the resulting output states.
It allows the computation of any multi-time correlation function of the system by inserting control operations at the respective times.
The PT exists for \emph{any} environment and is also called the quantum comb~\cite{Chiribella2008}, (generalized) influence functional~\cite{Jorgensen2019, Ye2021}, and process matrix~\cite{Oreshkov2012}.
Generically it is a high rank tensor that grows exponentially with the number of time steps.
In many cases, however, it is possible to systematically discard negligible correlations and express the PT as a tensor network~\cite{Orus2014} in a matrix product operator (MPO) form~\cite{Verstraete2004}, allowing a numerically efficient representation.
The necessary bond dimension of a PT-MPO reflects the degree of non-Markovianity in the interaction~\cite{Pollock2018a}.
For different environments different methods for the construction of a PT-MPO exist.
For linearly coupled Gaussian bosonic environments one can directly construct a tensor network that yields a PT-MPO~\cite{Jorgensen2019, Strathearn2019, Fux2021, OQuPy2022}.
Other approaches~\cite{Cygorek2021, Ye2021} allow the construction of PT-MPOs for any environment that can be approximated by a finite set of independent degrees of freedom.
It is also possible to construct PT-MPOs directly from experimental measurements~\cite{White2021}.

\begin{figure}[tb]
	\includegraphics[width=0.49\textwidth]{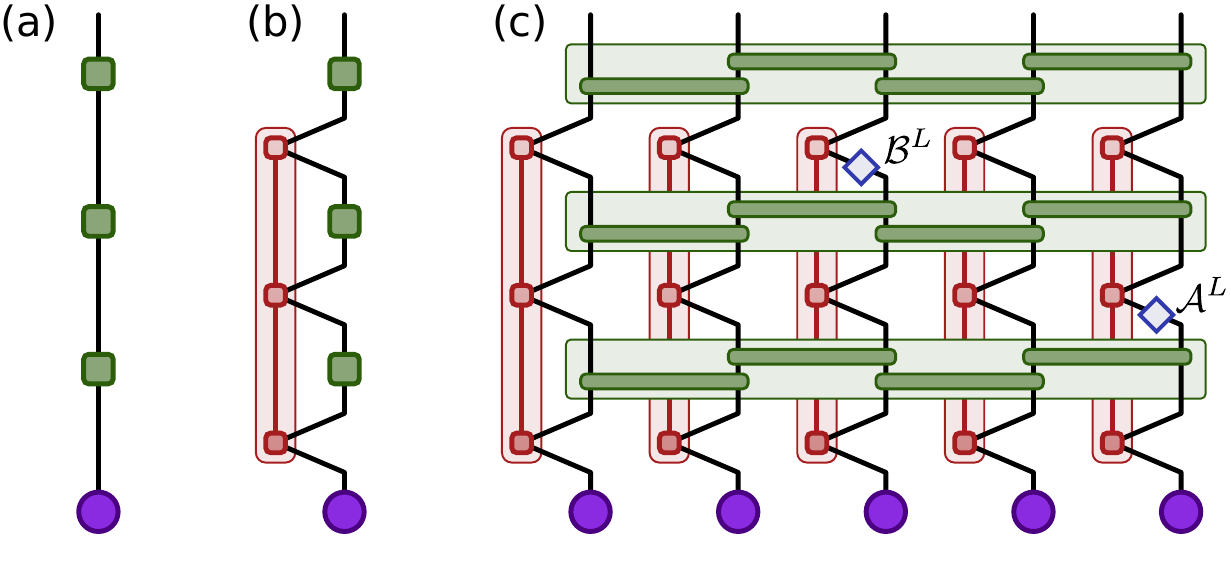}
	\caption{\label{fig:tensor-network-without-chain}%
		Tensor networks for the simulation of a single closed~(a), a single open~(b), and a chain of open quantum systems~(c).
		The purple circles represent the initial system states. The green squares and green rectangles represent the system propagators.
		The red and green shaded areas highlight the PT-MPOs and the TEBD propagators, respectively.
		In panel~(c) the super-operators $\mathcal{A}^L$ and $\mathcal{B}^L$ are inserted to calculate the two-time correlation \mbox{$\expect{\op{B}_3(2\,\Dt) \op{A}_5(1\,\Dt)}$}.
  }
\end{figure}%

The tensor network we propose for a chain of open quantum systems is presented in Fig.~\ref{fig:tensor-network-without-chain}c and can best be understood by first considering the simpler tensor networks for a single closed (Fig.~\ref{fig:tensor-network-without-chain}a) and a single open quantum system (Fig.~\ref{fig:tensor-network-without-chain}b).
The purple circle with one leg in Fig.~\ref{fig:tensor-network-without-chain}a represents the vectorized initial density matrix of the system.
The green boxes with two legs represent the propagator matrices $\mathcal{U}^\mathrm{S} = \exp\left(\mathcal{L}^\mathrm{S} \Dt \right)$ in Liouville space for a short \mbox{time step $\Dt$}, where $\mathcal{L}^\mathrm{S} = -i [\op{H}^{S}, \cdot]$ is the system Liouvillian associated with the system \mbox{Hamiltonian $\op{H}^{S}$}.
The entire diagram in Fig.~\ref{fig:tensor-network-without-chain}a has one unconnected leg and thus represents a vector, namely the vectorized system density matrix after three time steps.

Figure~\ref{fig:tensor-network-without-chain}b shows the tensor network for the evolution of a single \emph{open} quantum system.
The red shaded region contains a PT-MPO obtained by one of the methods mentioned above.
This tensor network relies on a Suzuki-Trotter expansion of the total \mbox{propagator $e^{-i \op{H} \Dt}$} into propagators $e^{-i \op{H}^{S} \Dt} \: e^{-i \op{H}^{E} \Dt}$ of the pure \mbox{system part $\op{H}^{S}$} and the \mbox{remainder $\op{H}^{E}$}, plus higher order \mbox{terms $\mathcal{O}(\Dt^2)$}.
The time step $\Dt$ must be chosen small enough such that these higher-order terms can be neglected.
The entire tensor network represents the vectorized system density matrix after three time steps taking the environment influence into account.

Figure~\ref{fig:tensor-network-without-chain}c shows the proposed tensor network for the simulation of a chain of system-environment pairs, which is a combination of the network shown in Fig.~\ref{fig:tensor-network-without-chain}b with TEBD in Liouville space~\cite{Daley2004, Zwolak2004}.
We assume a total Hamiltonian of the form
\begin{equation}
	\op{H} = \sum_{n=1}^{N} \left( \op{H}^\mathrm{S}_n + \op{H}^\mathrm{E}_n \right) + \sum_{n=1}^{N-1} \op{K}_{n,n+1} \mathrm{,}
\end{equation}
with $N$ system-environment pairs and nearest neighbor couplings $\op{K}_{n,n+1}$ among the systems.
Note that we assume that each environment is coupled only to one system site.
Models in which environments simultaneously couple to multiple sites or directly among each other are outside the scope of this work.
For each system site, the effects of interactions with its environment can thus be encoded in separate PT-MPOs (the red shaded areas).
Because the chain interacts repeatedly with the environment (at multiple times), it is important to have a correct encoding of multi-time correlations as captured by the process tensor.
The green shaded areas represent the propagation of the closed system chain for short time steps $\Dt$.
For ease of presentation Fig.~\ref{fig:tensor-network-without-chain}c shows the tensor network for the propagation of the closed chain in a first-order Suzuki-Trotter splitting among the chain sites, but higher order expansions are also possible.
Here, the time step needs to be chosen small enough such that the Suzuki-Trotter expansion of the evolution is valid for both the system-environment and the internal system-system coupling terms.
Also for ease of presentation, the tensor network shown in Fig.~\ref{fig:tensor-network-without-chain}c is restricted to uncorrelated initial states.
We present the full tensor network for a second-order Suzuki-Trotter expansion and correlated initial states in appendix~\ref{app:PT-TEBD}.
For a Markovian environment (for which the internal legs of the process tensor disappear~\cite{Pollock2018a}) this tensor network reduces to a TEBD network in Liouville space.

This method is also related to an approach introduced by Ba\~{n}uls \textit{et al.}~\cite{Banuls2009, Muller-Hermes2012} to study a subsystem of an infinite chain by contracting a conventional TEBD network in the spatial direction, which can be understood as the construction of a PT-MPO.
Turning back to Fig.~\ref{fig:tensor-network-without-chain}c, the two additional blue diamond shaped matrices show how a tensor network of this form can be used to extract multi-time correlations such as $\expect{\op{B}_m(t_2) \op{A}_n(t_1)}$ for arbitrary system operators $\op{A}$, $\op{B}$ at sites $n$, $m$, and times $t_1$, $t_2$ respectively.
For this, we multiply the system propagators with the left acting super-operators $\mathcal{A}^L(\rho) = \op{A} \rho$ at time step $t_1$ and $\mathcal{B}^L$ at time step $t_2$.
We provide this method as a part of the open source python package OQuPy~\cite{OQuPy2022}.

\subsection{Scaling and comparison to other approaches}
The PT-MPO approach is limited to chains whose state can be well approximated by an MPS of some finite bond dimension $\chi$, as well as environments whose process tensor can be well approximated by an MPO of bond dimension $\xi$.
The computational complexity is then dominated by performing the singular value decompositions (SVD) involved in compressing the spatial MPS after the application of the system propagators.
In the worst case the largest matrices involved are of the dimension $(\chi \xi d^2) \times (\chi \xi d^2)$, where $d$ is the Hilbert space dimension of a single site.
We suggest a contraction and SVD sequence that reduces this dimension to $(\eta d^2) \times (\eta d^2)$ with $\chi \lesssim \eta \leq \chi\xi$ in appendix~\ref{app:PT-TEBD}.
The overall simulation of an $N$ site chain for $K$ time steps thus takes $\mathcal{O}(N K \eta^3 d^6)$ operations.
This algorithm is, like the canonical TEBD algorithm, well suited for parallel computing, since each pair of neighboring sites can be evolved separately.

We note that alternative numerical approaches to compute the dynamics of chains of sites coupled to individual environments exist~\cite{Prior2010, Chin2010, Makri2018a, Makri2018, Suzuki2019, Mascherpa2019, Purkayastha2020, Makri2021, Kundu2021, Kundu2021a, Flannigan2021, Bose2021}.
A method proposed by Suzuki \textit{et al.}~\cite{Suzuki2019} is based on the transfer matrix approach and restricted to Gaussian bosonic environments as well as diagonal system-system couplings (with respect to the local system basis).
The modular path integral (MPI) method \cite{Makri2018} was originally based on the same assumptions, but has recently been extended to more general cases~\cite{Kundu2021, Kundu2021a}.
Another approach is based on a quantum state diffusion method, also assuming Gaussian bosonic environments \cite{Flannigan2021}.
Recently, Bose and Walters proposed a multi-site decomposition of the tensor network path integral (MS-TNPI)~\cite{Bose2021}, which is similar to the method presented in this Paper, but again restricted to Gaussian bosonic environments and comparatively short memory times (only 4 time steps are presented).
In the special case where only the end sites couple to environments, methods such as Time Evolving Density matrices using Orthogonal Polynomials (TEDOPA) can be used, where the baths are mapped to extended chains~\cite{Prior2010, Chin2010, Purkayastha2020}.

These alternative approaches attempt to tackle the numerical complexity of both the system-system and the system-environment correlations simultaneously.
In contrast to that, the PT-MPO approach tackles these challenges sequentially by first systematically reducing the numerical complexity originating from system-environment correlations before integrating them into the full many-body problem.
Furthermore, the PT-MPO approach is not restricted to bosonic environments or short memory times~\cite{Cygorek2021,Ye2021}.
Altogether, this enables us to tackle a much broader class of problems.


\section{Equilibriation of an XYZ spin chain with thermal leads}
\label{sec:xyz}
To illustrate the use of this method we now turn to the study of an XYZ spin chain with strongly coupled thermal leads.
We also use this example to demonstrate a general approach, employing the FDT, to study the thermalization of subsystems even when the coupling to their environment is strong.
We consider the chain Hamiltonian
\begin{equation}\label{eq:xyz-spin-chain}
	\op{H}_\mathrm{XYZ} =  \sum_{n=1}^{N} \epsilon_n \op{s}^z_n + \sum_{n=1}^{N-1}  \sum_{\gamma\in\{x, y, z\}} J^\gamma \: \op{s}^\gamma_n \op{s}^\gamma_{n+1} \mathrm{,}
\end{equation}
where $\op{s}^\gamma_n = \op{\sigma}^\gamma_n/2$ denote the spin-1/2 operators at site~$n$.
We choose $J^x = 1.3$, $J^y = 0.7$, and $J^z = 1.2$ to break the symmetries of the Heisenberg model, and start with homogeneous on-site energies $\epsilon_n=1$.
We set $\hbar=k_\mathrm{B}=1$ and express all frequencies and times in units of some characteristic frequency and its inverse.

\subsection{Spin chain coupled to a single bath} 
As a first check we couple a single bath at temperature $T=1.6$ to only the first site of a short ($N=5$) chain, as sketched in Fig.~\ref{fig:spin-chain-N5-z3-with-chain}e.
We aim to confirm that at the steady state each spin has come to thermal equilibrium at temperature $T$.
The bath couples to the chain through an operator on the first site (which we choose to be $\op{s}^y_1$) with
\begin{equation}\label{eq:bosonic-bath}
	\op{H}^\mathrm{E}_1 =\sum_{k=0}^{\infty} \left[ \op{s}_1^y \left( g_k \op{b}_k^\dagger + g_k^* \op{b}_k \right) + \omega_k \op{b}_k^\dagger \op{b}_k \right] \mathrm{,}
\end{equation}
where $\op{b}_k^{(\dagger)}$ are bosonic lowering (raising) bath operators.
The $g_k$ parameters are determined by the spectral density  \mbox{$J(\omega)=\sum_k \abs{g_k}^2 \delta(\omega-\omega_k)$}, which we choose to take the form
\mbox{$J(\omega)=2\alpha\omega \exp(-\omega^2/\omega^2_c)$}.
We use a coupling strength \mbox{$\alpha=0.32$} and cutoff frequency \mbox{$\omega_c=4.0$}, which are of the same order of magnitude as the parameters for a quantum dot exciton interacting with its phonon environment given in units of $\mathrm{ps}^{-1}$~\cite{Ramsay2010}.

\subsubsection{Comparison to Gibbs state in the weak coupling limit}
\begin{figure}[t]
	\includegraphics[width=0.49\textwidth]{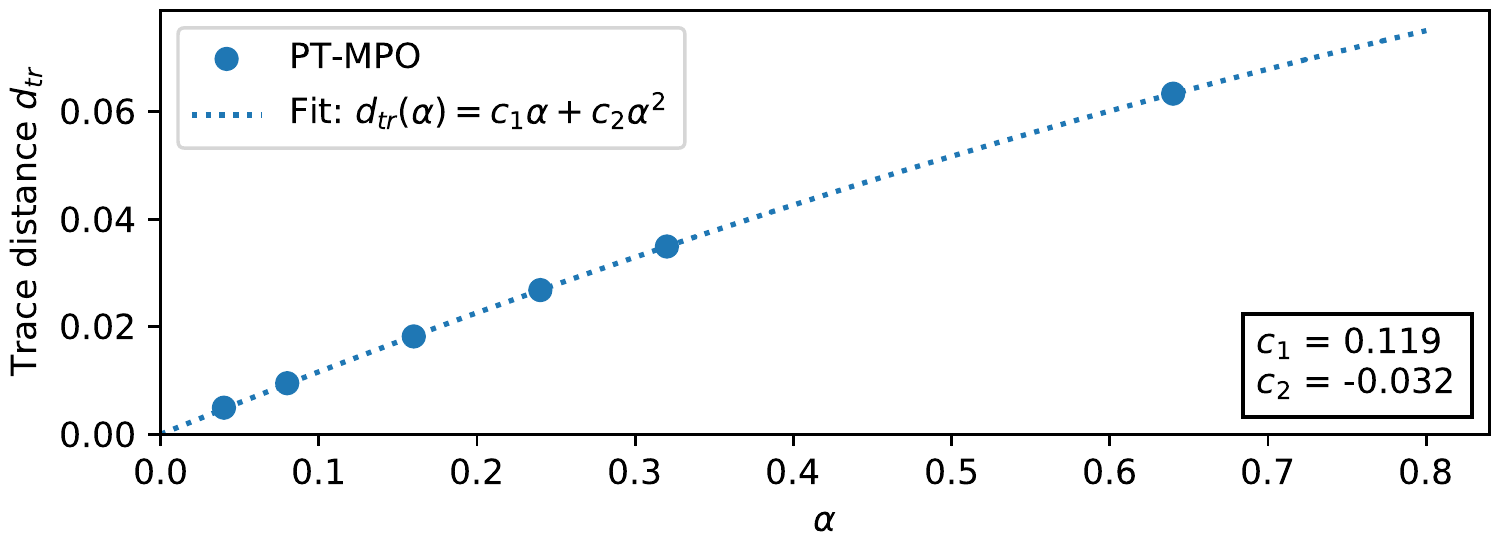}
	\caption{\label{fig:weak-coupling}%
		Trace distance between the thermal Gibbs state of the closed 5-site spin chain and the approximate steady state of the chain coupled to the bath with varying coupling strength $\alpha$.
		A fit shows that this difference is vanishing at a linear order in $\alpha$ (i.e. a quadratic order in bath couplings $g_k$) in a weak coupling limit.}
\end{figure}%
Perturbation theory predicts that, in a weak coupling limit, the reduced chain density matrix of the full thermal state differs from the Gibbs state of the chain Hamiltonian at a quadratic order in the bath coupling~\cite{Trushechkin2021}.
The dimensionless coupling strength $\alpha$ is proportional to the square of the bath coupling amplitudes $\abs{g_k}^2$, i.e. $\alpha \propto \sum_k\abs{g_k}^2$.
Assuming that the chain thermalizes with the bath in the long time limit, we thus expect to find a difference between the reduced steady state and the Gibbs state of the chain Hamiltonian that is proportional to $\alpha$ in a weak coupling limit.
Figure~\ref{fig:weak-coupling} shows the trace distance for various coupling strengths $\alpha$.
A fit to the data shows that the results are consistent with the expectation.

While it is reassuring that this method leads to known results in a weak coupling limit, it seems desirable to validate that the spin chain has thermalized for finite coupling strengths $\alpha$.
The exact reduced thermal chain state in such cases is, however, unknown and we thus lack a reliable reference~\cite{Trushechkin2021}.
As mentioned in the introduction, beyond the reduced density matrix there is additional information encoded in multi-time correlations.
We can use this information to validate that the chain and environment have indeed thermalized by checking the consistency of two-time correlations with the FDT.

\subsubsection{Checking thermalization through the fluctuation--dissipation theorem}
The FDT~\cite{Kubo1966, Kubo1978} states that for a thermalized quantum system at \mbox{temperature $T$} the ratio of the fluctuation and dissipation spectra with respect to any observable $\op{A}$ must be 
\begin{equation}
    \frac{S_{A}(\omega)}{\chi''_{A}(\omega)} = \coth\left( \frac{\omega}{2 T} \right).
\end{equation}
Here, the fluctuation spectrum $S_{A}(\omega)$, also known as the symmetrized quantum noise spectral density~\cite{Clerk2010}, is the Fourier transform of the Keldysh Green's function \mbox{$S_{A}(\tau)=\frac{1}{2}\expect{\{\op{A}(\tau),\op{A}(0)\}}$}.
Similarly, the dissipation spectrum $\chi''_{A}(\omega)$ is the imaginary part of the Fourier transformed linear response function \mbox{$\chi_{A}(\tau)=i\Theta(\tau)\expect{[\op{A}(\tau),\op{A}(0)]}$} and quantifies the density of states for transitions driven by the operator $\op{A}$.

The validity of the FDT is a general and exact result of statistical quantum mechanics and does not involve any weak-coupling approximations.
Weak-coupling (i.e. linear response theory) is however generally invoked in considering how one could measure these two-time correlations in an experiment.
This can be done by weakly coupling a measurement device to the degree of freedom $\op{A}$ of the system and recording its fluctuation and dissipation spectra.
As such, the temperature found from checking the FDT for a particular observable corresponds to weakly coupling a thermometer to that part of the system.
Even here, although we assume $\op{A}$ couples weakly to the measurement device, it may still couple strongly to other parts of the system and environment.

\begin{figure}[t]
	\includegraphics[width=0.49\textwidth]{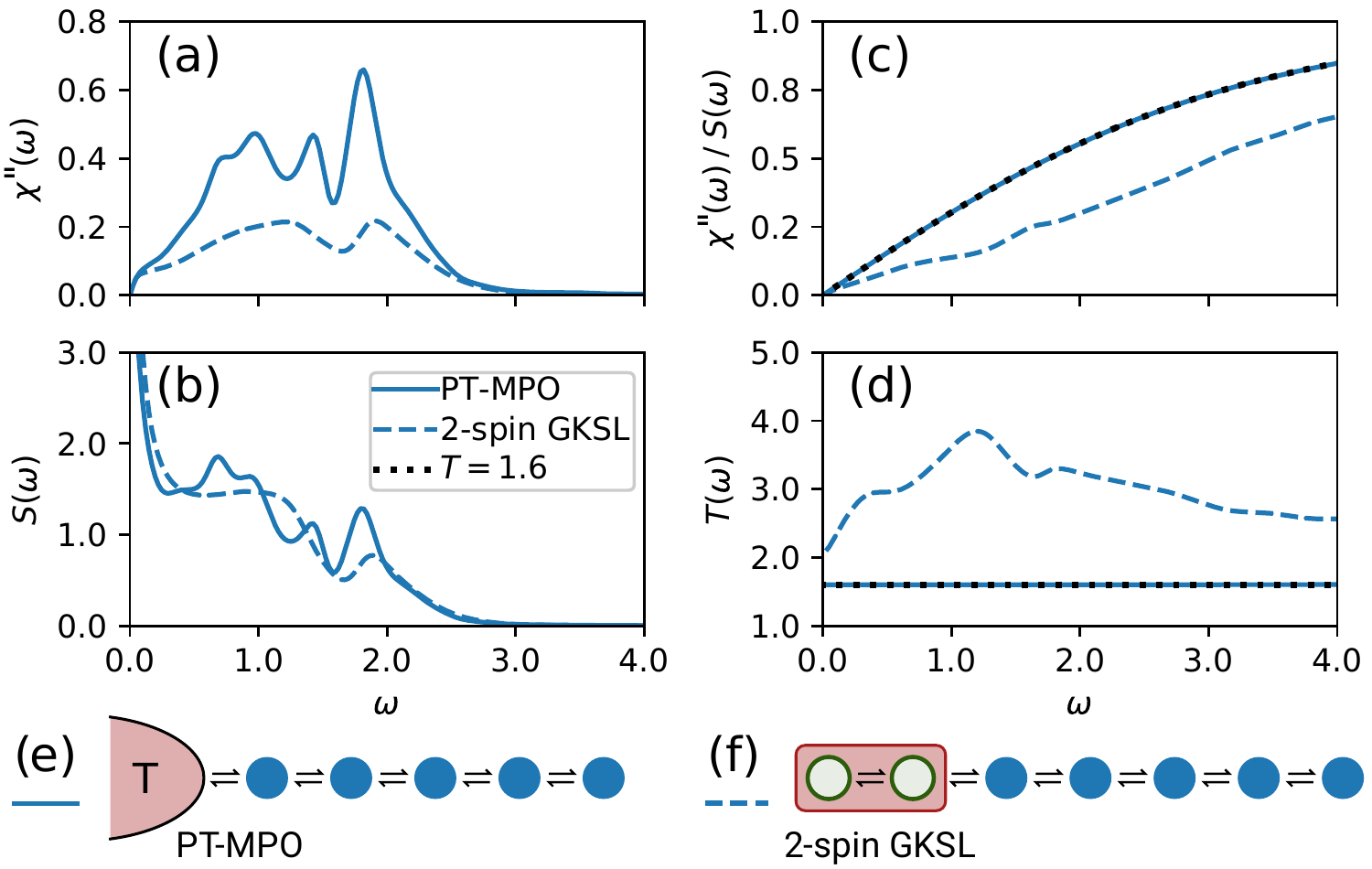}
	\caption{\label{fig:spin-chain-N5-z3-with-chain}%
		Two-time correlations of the steady state for the $\op{\sigma}^z$ observable of the middle spin in a 5-site chain coupled to a single bath at temperature $T=1.6$.
		The solid and dashed lines show the results obtained employing the PT-MPO approach and a 2-spin driving protocol~\cite{Prosen2009, Znidaric2010, Prosen2011}, respectively.
		The panels show the dissipation spectrum~(a), the fluctuation spectrum~(b), their ratio~(c), and the corresponding effective temperature~(d).
		The PT-MPO results  overlap with the expected FDT $\mathrm{tanh}(\frac{\omega}{2T})$ (dotted line) in~(c) and show no frequency dependence in~(d), confirming complete thermalization.
		Panel~(e) shows a sketch of the 5-site spin chain coupled to a single bath.
            Panel~(f) shows a sketch of a model in which the bath has been replaced by two additional spins, which are driven to their local thermal Gibbs state with a GKSL master equation.
        }
\end{figure}%
Figure~\ref{fig:spin-chain-N5-z3-with-chain} shows the simulation results for the fluctuation and dissipation spectra for the $\op{\sigma}^z$ observable of the middle spin of the 5-site chain with a system-environment coupling strength of $\alpha=0.32$.
The solid lines show the results obtained using the PT-MPO technique (for the details of the numerical simulation see appendix~\ref{app:XYZ-details}).
Figure~\ref{fig:spin-chain-N5-z3-with-chain}c shows the ratio of the dissipation and fluctuation spectra, which has the shape of a hyperbolic tangent.
Inverting the FDT and plotting a frequency dependent effective temperature,
\begin{equation}
T(\omega) = \frac{\omega}{2\, \mathrm{artanh}[\chi''(\omega)/S(\omega) ]},
\end{equation}
in Fig.~\ref{fig:spin-chain-N5-z3-with-chain}d we can see from the perfectly flat line that the two-time correlations are consistent with the FDT at the expected temperature.
We find similar results for all other spins and observables, and no dependency on the chosen initial chain state.

\subsubsection{Comparison to two-spin GKSL driving}

Figure~\ref{fig:spin-chain-N5-z3-with-chain} also shows the results of a different, widely applied numerical method to study thermodynamic properties of spin chains~\cite{Prosen2009, Znidaric2010, Prosen2011}.
In this approach, two additional spins are attached to the end of the chain and driven towards their local two-spin thermal Gibbs state with a time local master equation of GKSL form in the hope that this will thermalize the rest of the chain.

To compute the dashed lines in Fig.~\ref{fig:spin-chain-N5-z3-with-chain} we employed the two-spin bath protocol introduced in \cite{Prosen2009}.
For this we attach two additional spins (at positions $n=-1$ and $n=0$) to the left hand side of the first spin and construct a Liouvillian $\LL_B$ which drives these two spins towards the Gibbs state of their local Hamiltonian as described in section 2.4 of reference \cite{Prosen2009}.
Because the augmented TEBD method reduces to the canonical TEBD in Liouville space when no PT-MPOs are added to the network, we can use the exact same approach and implementation as described above.
For this, we simply do not attach any PT-MPO to the network, and instead substitute $\LL^{K^\prime}_{-1,0} \rightarrow \LL^{K^\prime}_{-1,0} + \LL_B$ to include the time local driving terms.

As can be seen from the figure, the two-time correlations obtained using this method strongly deviate from the FDT and are thus incorrect.

\subsection{Effective temperature in a thermal gradient}

\begin{figure}[tb]
	\includegraphics[width=0.49\textwidth]{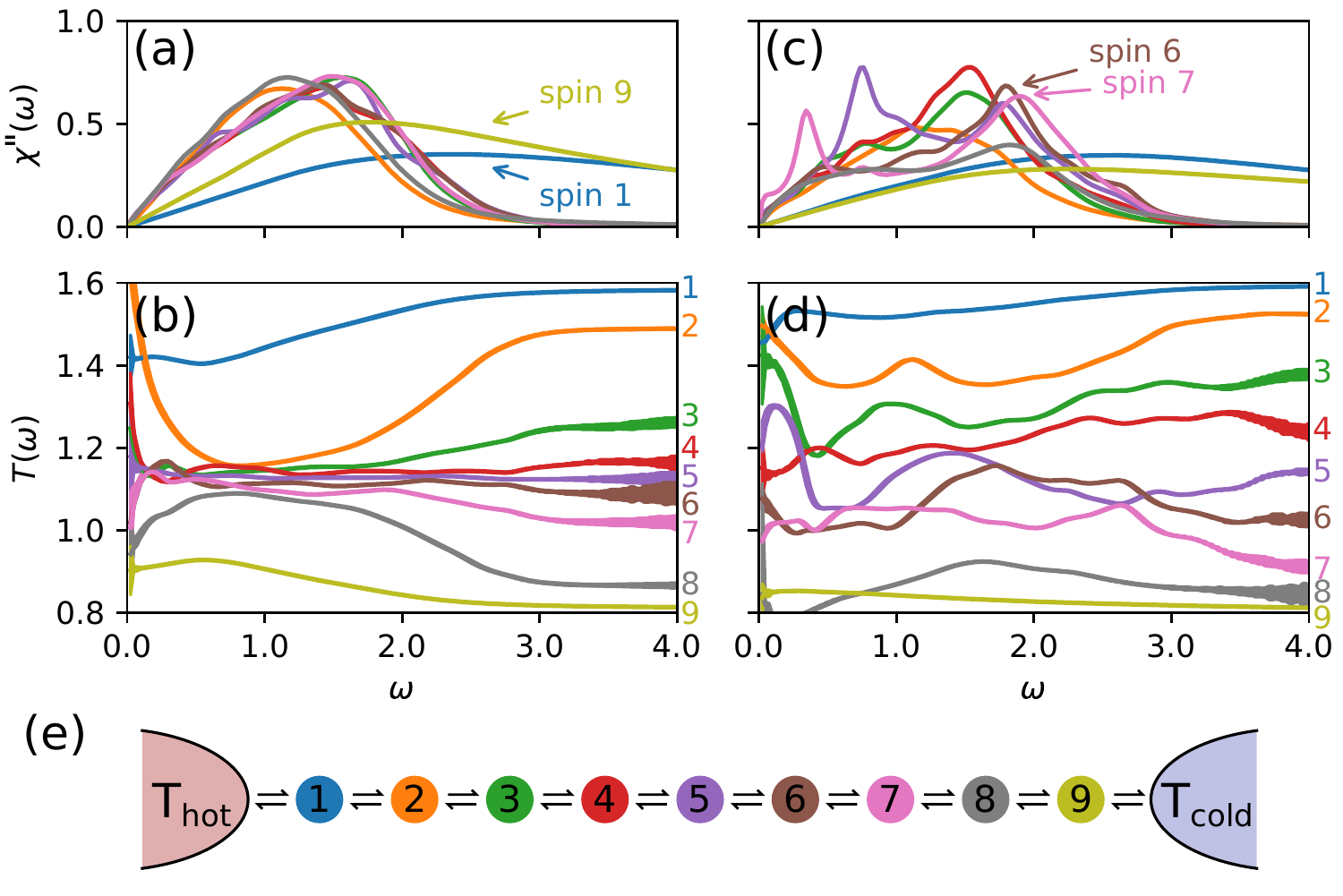}
	\caption{\label{fig:spin-chain-N9-z-with-chain}%
		The dissipation spectrum~(a) and effective temperature~(b) at steady state for the $\sigma^z$ observable of each spin in a 9-site spin chain placed between a hot ($T_\mathrm{hot}=1.6$) and cold bath ($T_\mathrm{cold}=0.8$).
		Panels (c) and (d) show the results for the same chain with additional on-site disorder \mbox{$\epsilon_n = 1 + x_n$}, for a random draw of $x_n$ from a uniform distribution in $(-1.6, 1.6)$.
		The thickness of the lines in panel (b) and (d) represent an estimate of the numerical error.
		Panel~(e) shows a sketch of the 9-site spin chain coupled to two thermal baths at different temperatures, and serves as a legend for the line colors used in the other panels.
	}
\end{figure}

Having demonstrated the expected thermalization for a single bath, we now turn to an XYZ spin chain of length $N=9$ coupled to two thermal leads at different temperatures, as sketched in Fig.~\ref{fig:spin-chain-N9-z-with-chain}e. Thermalization in this larger system would be challenging to address with other methods.
Using the PT-MPO method we couple one bath at temperature $T_\mathrm{hot}=1.6$ to the first spin and one bath at $T_\mathrm{cold}=0.8$ to the last spin (see Fig.~\ref{fig:spin-chain-N9-z-with-chain}e), using the same spectral density and coupling operator $\op{s}^y$ as before.
Figures~\ref{fig:spin-chain-N9-z-with-chain}a~and~\ref{fig:spin-chain-N9-z-with-chain}b show the dissipation spectrum $\chi''(\omega)$ and the effective temperature $T(\omega)$ for the $\op{\sigma}^z$ observable of each spin.

In Fig.~\ref{fig:spin-chain-N9-z-with-chain}b we observe that at a mid frequency range (between roughly $0.5$ and $2.0$) the inner spins adopt a common intermediate effective temperature, while at higher frequencies (above approximately $3.0$) each spin adopts an effective temperature between that of the hot and cold bath depending on its position.
In the following we suggest an idea for why this kind of behavior might arise.

\begin{figure}[tb]
	\includegraphics[width=0.49\textwidth]{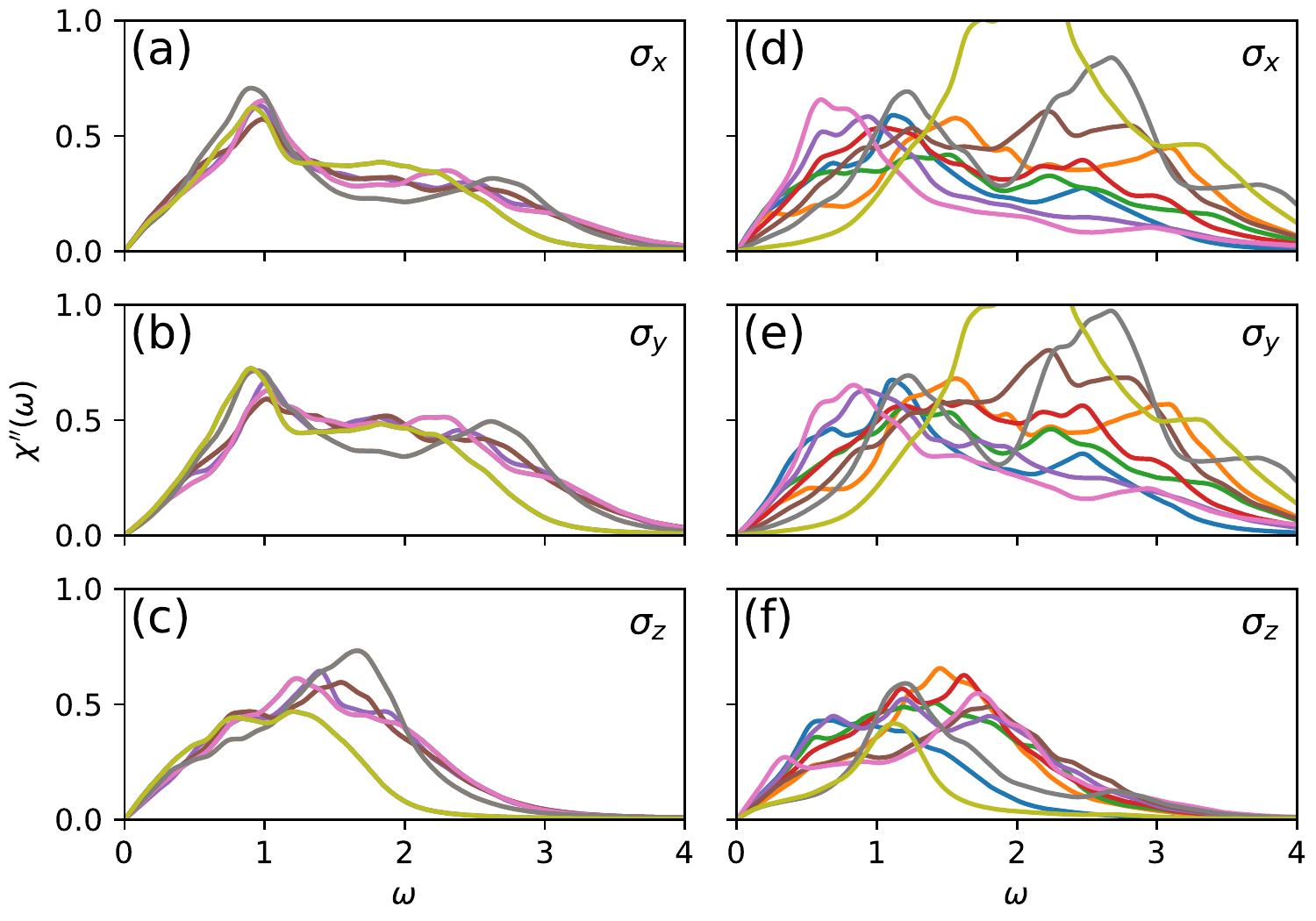}
	\caption{\label{fig:density-of-states-N9}%
		Density of states $\chi''(\omega)$ for the closed 9-site spin chain with respect to $\op{\sigma}^x_n$, $\op{\sigma}^y_n$, and $\op{\sigma}^z_n$ for each site $n$, and the clean (a-c) and disordered case (d-f).
        We obtained these results from exact diagonalization and included a line broadening with a Lorentzian shape for each mode (with a width of $0.1$) in order to approximate the effect of the environment.  The color of lines corresponding to different sites matches those shown in Fig.~\ref{fig:spin-chain-N9-z-with-chain}
  }
\end{figure}

We first consider the eigenstates of the closed XYZ spin chain, which consist of a set of delocalized ``bulk'' states and localized ``surface'' states.
The surface states are mainly localized each at one end of the chain, but reach into the bulk with an exponentially decaying tail.
This can be shown quantitatively by plotting the local density of states on each site; this is shown in Fig.~\ref{fig:density-of-states-N9}.
For the chain parameters chosen here the density of states for the closed spin chain slowly vanishes above a frequency of approximately $2.5$ (see Fig.~\ref{fig:density-of-states-N9}c).
In the mid frequency range the density of states for the inner spins is dominated by the bulk states.
When we include the coupling to the environments, the bulk states hybridize weakly with both environments due to their equal and small overlap with the two outer spins, which leads to the intermediate common temperature of the inner spins.
For the higher frequencies, however, the density of states is dominated by the surface states.
This is because the surface states have a large overlap with either the first or last spin and thus hybridize strongly with the left or right environment respectively.
Because the coupling of a spin with the left and right surface states strongly depends on its position, the effective temperature it adopts depends on its position as well.
This picture is also consistent with the dissipation spectrum plotted in Fig.~\ref{fig:spin-chain-N9-z-with-chain}a, showing different behavior for the inner and outer spins.

Since the explanation suggested above depends on the difference between delocalized bulk states and localized surface states, it can be tested by adding disorder to localize the bulk states.  This is shown in Fig~\ref{fig:spin-chain-N9-z-with-chain}d. 
For all figures with disorder we use the following energies,
 \mbox{$\epsilon_n = 1 + x_n$} with \mbox{$x_n = (0.16, 0.69, 0.33, 0.14, -0.24, 0.47, -0.20, 1.25, 1.48)$}.
With the addition of disorder,  all sites become localized. We clearly see that this destroys the collective common temperature at mid frequencies as expected.

\subsubsection{Response functions for $\op{\sigma}^x$ and $\op{\sigma}^y$}
\begin{figure}[tb]
	\includegraphics[width=0.49\textwidth]{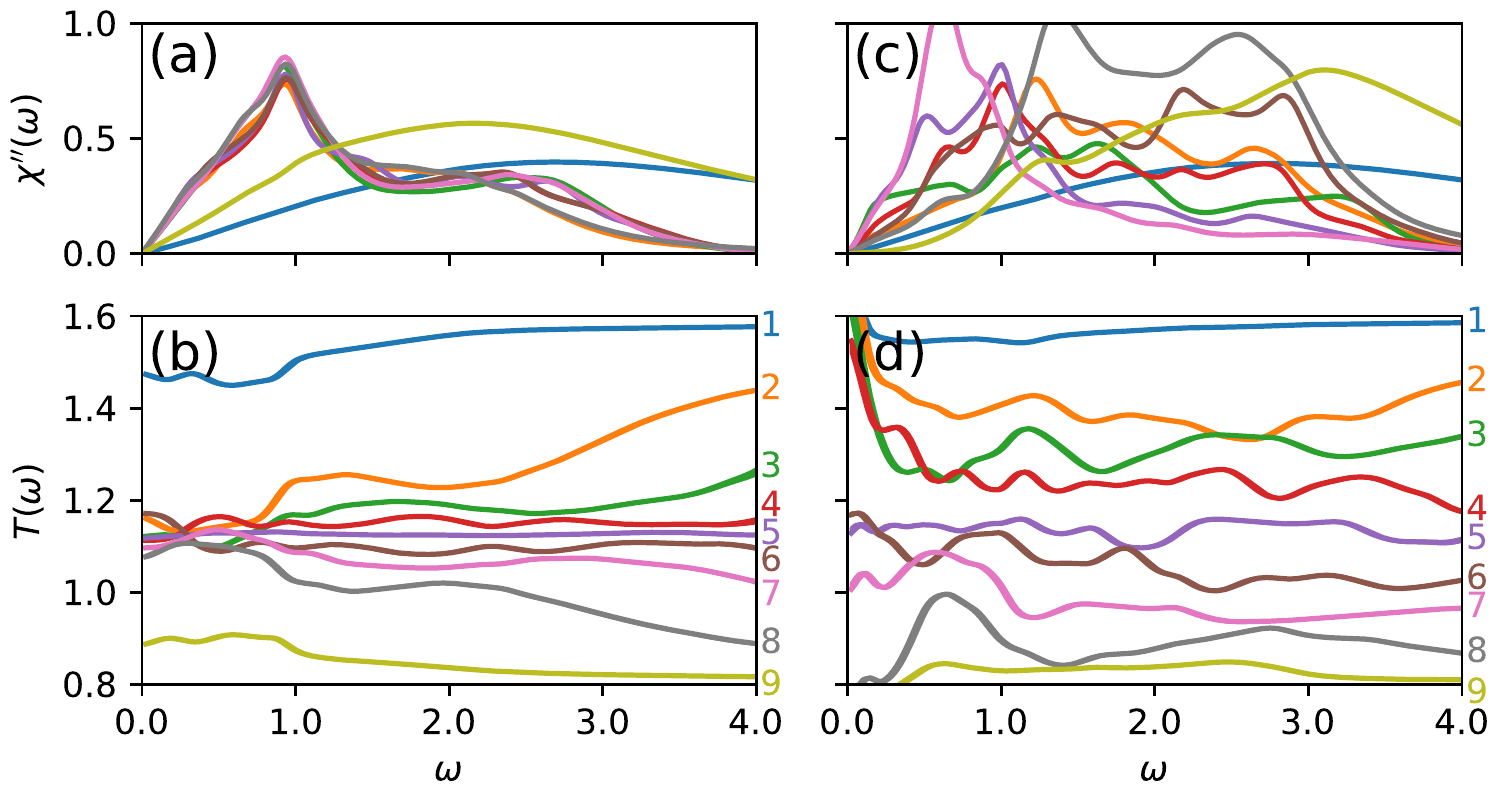}
	\caption{\label{fig:spin-chain-N9-x}%
		The dissipation spectrum~(a) and effective temperature~(b) at steady state for the $\op{\sigma}^x$ observable of each spin in a 9-site spin chain placed between a hot ($T_\mathrm{hot}=1.6$) and cold bath ($T_\mathrm{cold}=0.8$).
		Panels (c) and (d) show the results for the same chain with additional on-site disorder.}
\end{figure}
\begin{figure}[tb]
	\includegraphics[width=0.49\textwidth]{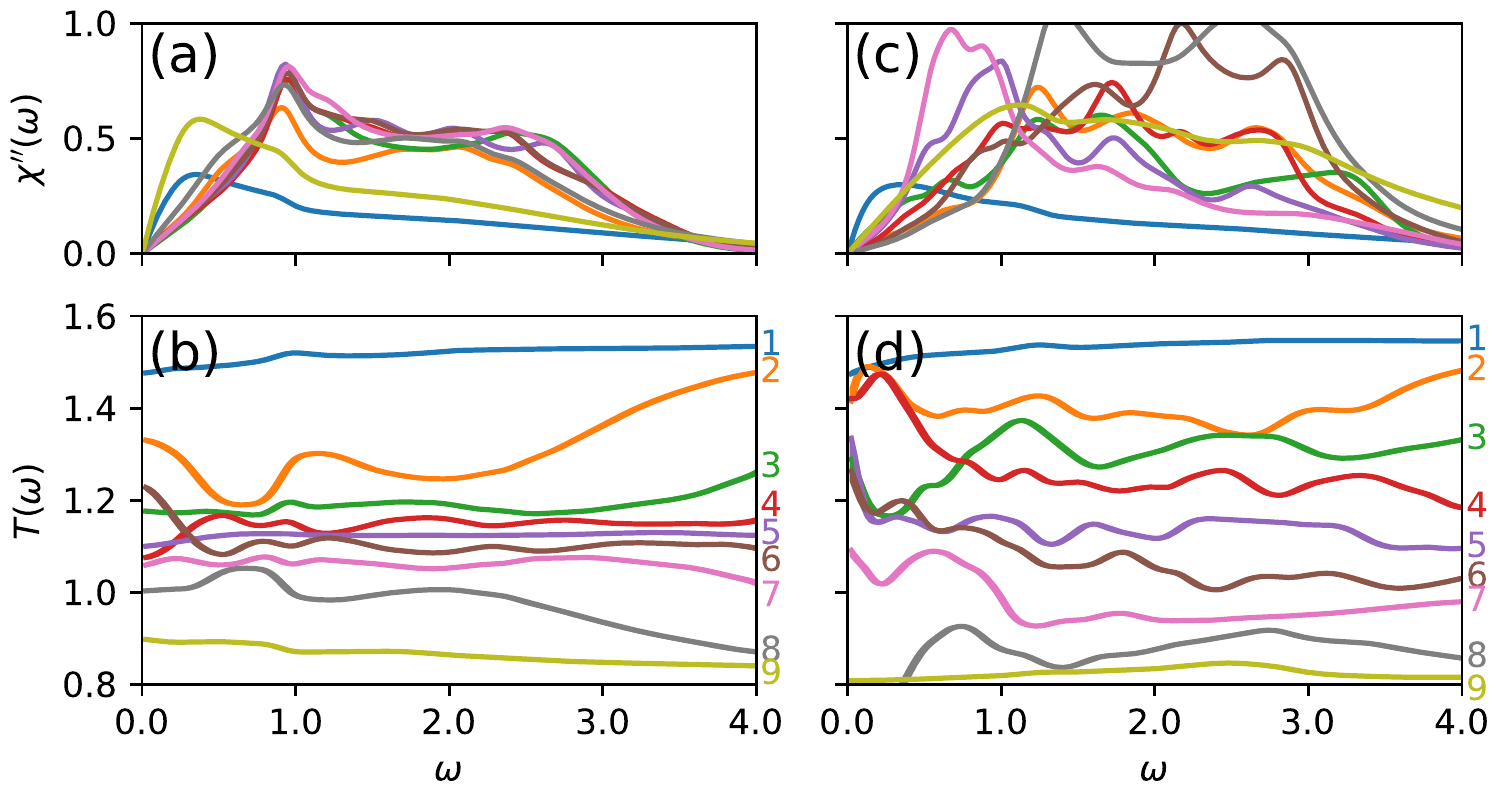}
	\caption{\label{fig:spin-chain-N9-y}%
		The dissipation spectrum~(a) and effective temperature~(b) at steady state for the $\op{\sigma}^y$ observable of each spin in a 9-site spin chain placed between a hot ($T_\mathrm{hot}=1.6$) and cold bath ($T_\mathrm{cold}=0.8$).
		Panels (c) and (d) show the results for the same chain with additional on-site disorder.}
\end{figure}%
Figures~\ref{fig:spin-chain-N9-x}~and~\ref{fig:spin-chain-N9-y} show the dissipation spectra and the effective temperature with respect to the observables $\op{\sigma}^x$ and $\op{\sigma}^y$, and thus complement Fig.~\ref{fig:spin-chain-N9-z-with-chain}.
In all three figures we used the same on-site disorder as described above.
The effective temperature plots in Figs.~\ref{fig:spin-chain-N9-x}b~and~\ref{fig:spin-chain-N9-y}b do not show the distinct low, mid, and high frequency regions as in Fig.~\ref{fig:spin-chain-N9-z-with-chain}b.
However, the results are still consistent with the explanation suggested above.
For the operators $\op{\sigma}^{x,y}$ we see the density of states extend to higher frequencies (see Figs.~\ref{fig:density-of-states-N9}a~and~\ref{fig:density-of-states-N9}b).
This is because these operators couple spaces with different values of $\sum_n \hat{s}^z_n$, and the energies of these states are split by the on-site field $\epsilon_n$.

For the clean chain (without disorder), the inner spins tend to assume a collective common temperature at frequencies where their density of states is larger than the density of states for the outer spins.
Conversely, at higher frequencies  where the density of states of  the outer spins is dominant, the surface states have an increased influence, and the effective temperature is more spread out.

\section{A 21-site anisotropic XY spin chain with individual environments}
\label{sec:xy}

\begin{figure}
	\includegraphics[width=0.49\textwidth]{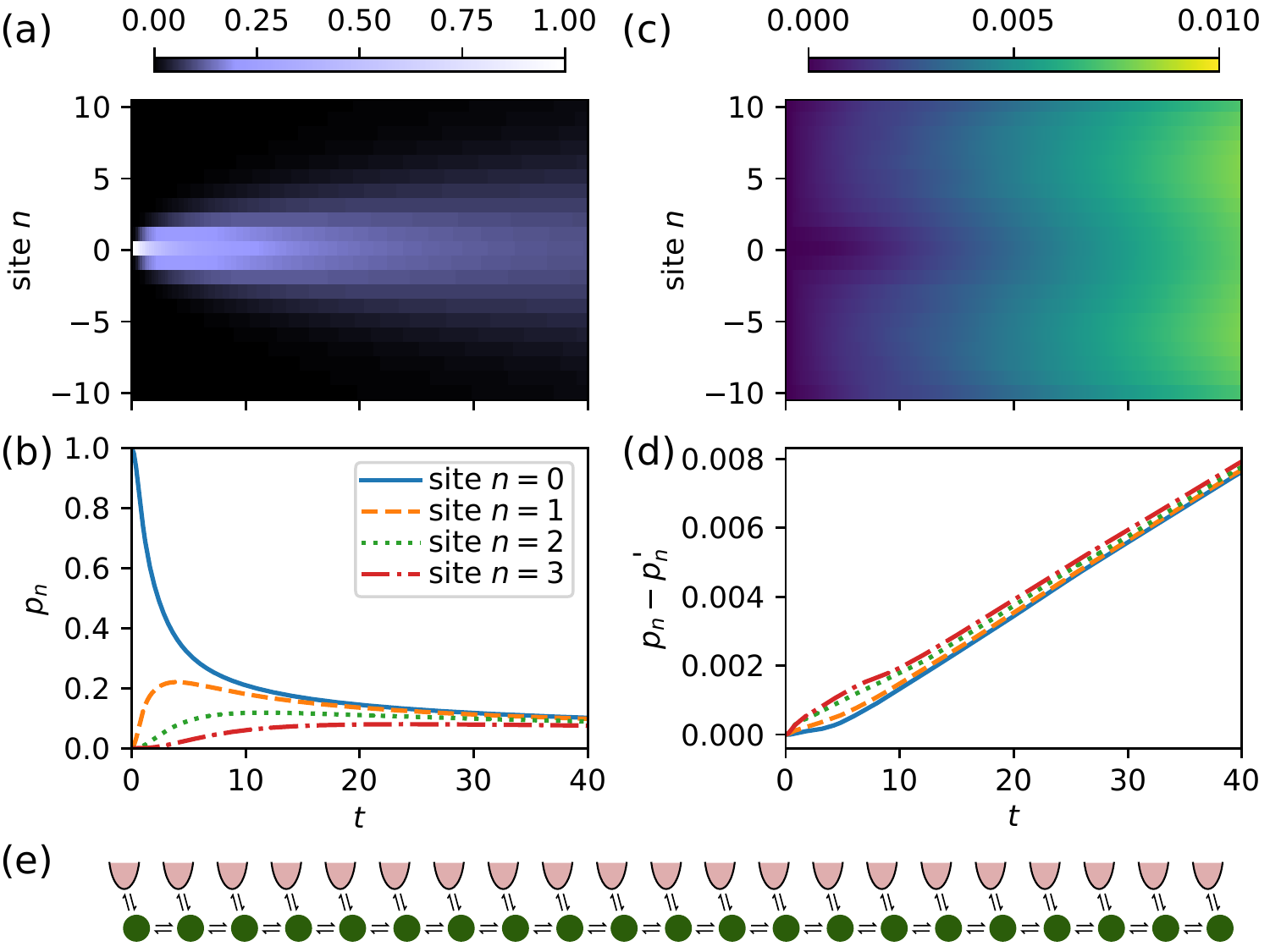}
	\caption{\label{fig:long-chain-a16-with-chain}%
		The dynamics of the XY spin chain with strongly coupled ($\alpha=0.16$) bosonic environments on each site starting from an initial excitation in the middle of the chain.
		The panels (a) and (b) show the dynamics of the excitation probability $p_n(t)$ of the spin at site $n$ for an XY interaction anisotropy of $\eta=0.04$.
		The panels (c) and (d) show the difference between $p_n(t)$ for the anisotropic case ($\eta=0.04$) and $p^\prime_n(t)$ for the isotropic case ($\eta=0.0$).
  		Panel~(e) shows a sketch of the 21-site spin chain where each spin couples to its individual environment.
	}
\end{figure}

To further demonstrate the power of our proposed method we study the dynamics of a 21-site (an)isotropic XY spin chain, for which each spin couples strongly to its own structured bosonic environment, as sketched in Fig.~\ref{fig:long-chain-a16-with-chain}e.
We study the spread of a single initial excitation in the middle of the chain for two different environment coupling strengths $\alpha \in \{0.16, 0.32\}$ and for a zero and non-zero XY coupling anisotropy $\eta \in \{ 0.0, 0.04 \}$.
The total Hamiltonian we consider is
\begin{equation}
	\op{H} = \op{H}_\mathrm{XY} + \sum_{n=-10}^{10} \op{H}^\mathrm{E}_n \, \mathrm{,}
\end{equation}
with the (an)isotropic XY chain Hamiltonian
\begin{equation}\label{eq:xy-spin-chain}
	\op{H}_\mathrm{XY} =  \!\!\sum_{n=-10}^{10} \!\op{s}^z_n
	+ \!\!\sum_{n=-10}^{9}\! \left[ (1\!-\!\eta) \op{s}^x_n \op{s}^x_{n+1} 
	+(1\!+\!\eta) \op{s}^y_n \op{s}^y_{n+1} \right] \mathrm{,}
\end{equation}
and the environment Hamiltonians
\begin{equation}\label{eq:bosonic-bath-XY}
	\op{H}^\mathrm{E}_n =\sum_{k=0}^{\infty} \left[ \op{s}_n^z \left( g_{n,k} \op{b}_{n,k}^\dagger + g_{n,k}^* \op{b}_k \right) + \omega_{n,k} \op{b}_{n,k}^\dagger \op{b}_{n,k} \right] \mathrm{,}
\end{equation}
where $\op{s}_n^\gamma$ are the spin-1/2 operators on site $n$, and $\op{b}_{n,k}^{(\dagger)}$ are bosonic lowering (raising) operators of the $k$-th mode of the $n$-th environment.
We again take the $g_{n,k}$ parameters in order to give an Ohmic spectral density, 
\mbox{$J_n(\omega)=2\alpha\omega \exp(-\omega^2/\omega_c^2)$} for all environments.
We choose a cutoff frequency \mbox{$\omega_c=4.0$} and assume that all environments are initially at thermal equilibrium at temperature $T=1.6$.
\begin{figure}
	\includegraphics[width=0.49\textwidth]{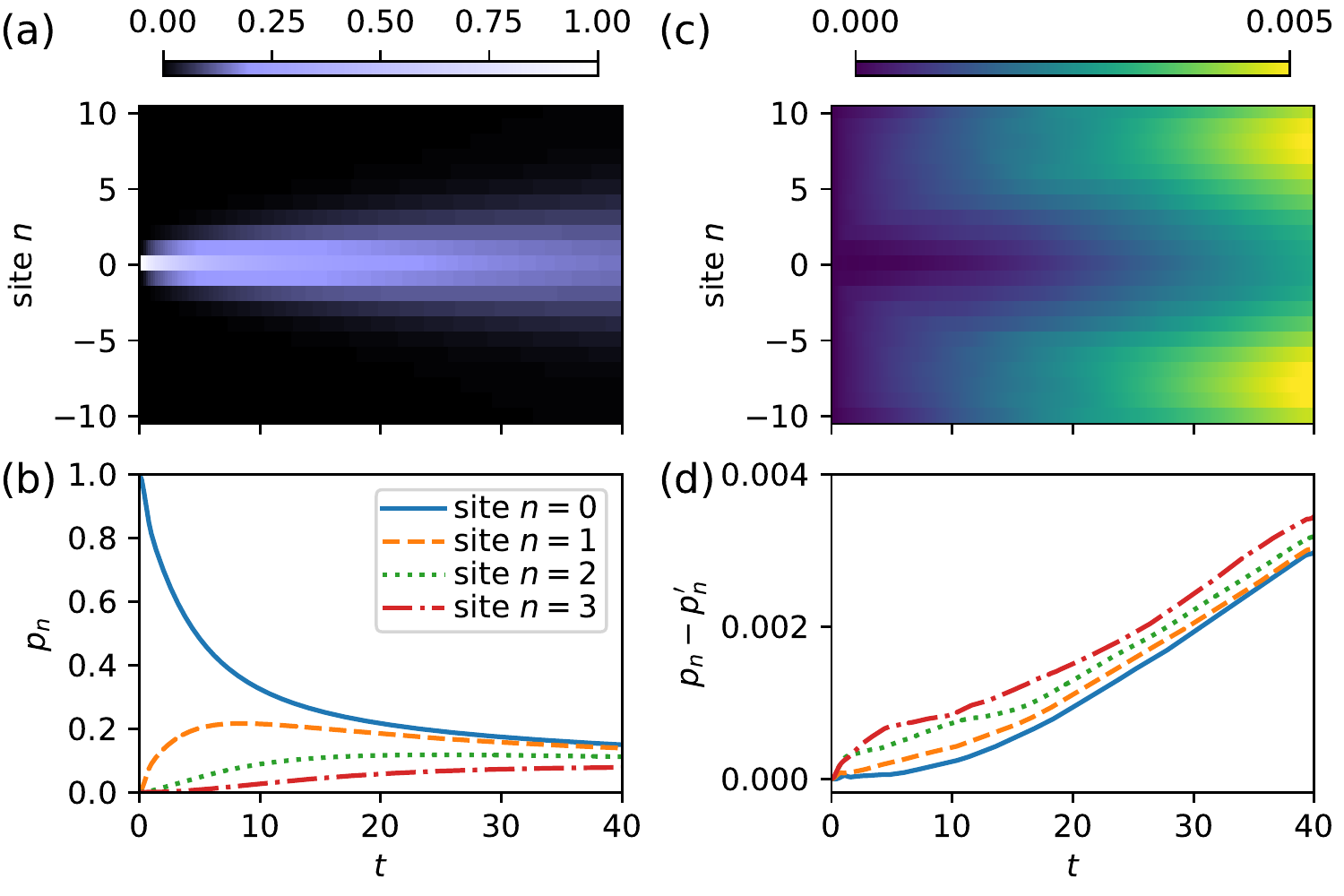}
	\caption{\label{fig:long-chain-a32}%
		The dynamics of the XY spin chain with strongly coupled ($\alpha=0.32$) bosonic environments on each site starting from an initial excitation in the middle of the chain.
		The panels~(a) and~(b) show the dynamics of the excitation probability $p_n(t)$ of the spin at site $n$ for an XY interaction anisotropy of $\eta=0.04$.
		The panels~(c) and~(d) show the difference between $p_n(t)$ for the anisotropic case ($\eta=0.04$) and $p^\prime_n(t)$ for the isotropic case ($\eta=0.0$).
	}
\end{figure}%

As a first step, we compute the PT-MPOs for each environment employing the PT-TEMPO method~\cite{Jorgensen2019, Strathearn2019, Fux2021} and its implementation in the open source Python package OQuPy~\cite{OQuPy2022}.
Because the environment Hamiltonians and temperature are identical for each environment we only need to compute one PT-MPO for each coupling strength $\alpha \in \{0.16, 0.32\}$ and can then augment each site of the TEBD tensor network with a separate copy.

Figures~\ref{fig:long-chain-a16-with-chain}~and~\ref{fig:long-chain-a32} show the chain dynamics for the environment coupling strengths $\alpha=0.16$ and $\alpha=0.32$, respectively.
The panels (a) and (b) each show the dynamics of the probability $p_n(t)$ to find the spin at site $n$ in the z-basis ``up'' state for the anisotropic case with $\eta=0.04$.
The panels (c) and (d) each show the difference between the anisotropic case $p_n(t)$ and the isotropic case $p^\prime_n(t)$.
In contrast to the isotropic case, the total spin excitation number is not preserved in the anisotropic case, for which we observe an approximately linear growth over time.

\begin{figure}
	\includegraphics[width=0.49\textwidth]{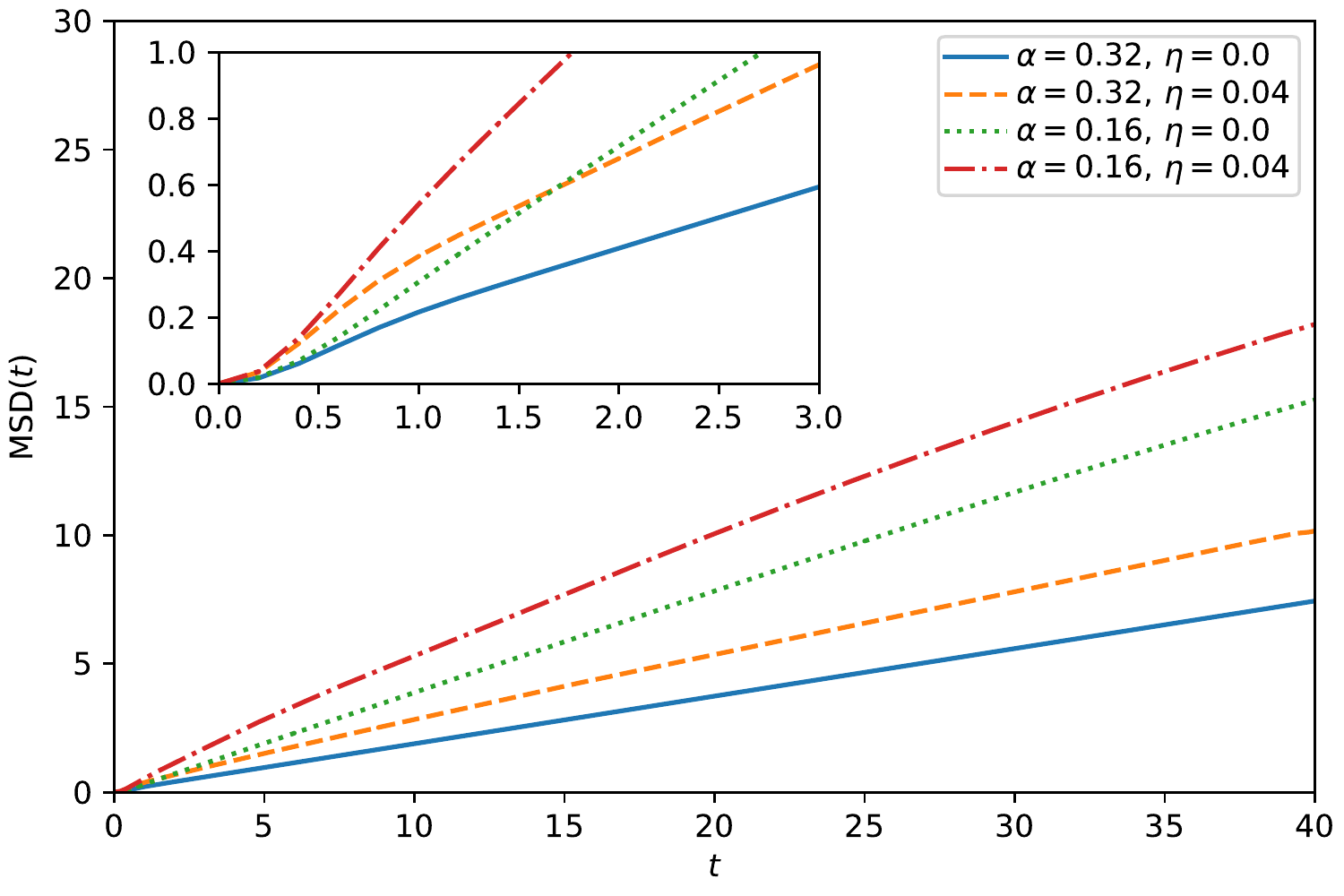}
	\caption{\label{fig:long-chain-displacement}%
		The mean-squared displacement (MSD) of an initial excitation in the XY spin chain with strongly coupled bosonic environments on each site for four combinations of environment coupling strength $\alpha$ and XY coupling anisotropy $\eta$.
		After some early time dynamics (shown in the inset), the MSD grows approximately linearly in time, suggesting a diffusive spread of the excitation.
		}
	
\end{figure}%

Finally, to help the interpretation of these results we plot the mean-squared displacement
\begin{equation}
	\mathrm{MSD}(t) = \frac{1}{p_n(t)} \sum_{n=-10}^{10} p_n(t) \times n^2
\end{equation}
in Fig.~\ref{fig:long-chain-displacement}.
The MSD appears to grow linearly at later times, i.e. it shows diffusive dynamics in all four cases.
The results suggest a larger diffusion constant for stronger anisotropies as well as for weaker environment coupling strengths.
For the very early-time dynamics, however, the MSD appears to be independent of the environment coupling strength.


\section{Conclusions}
We have presented a numerical method for computing multi-time correlations of many-body quantum systems in the presence of strongly coupled and structured environments and have demonstrated that two-time correlations can be used to study thermalization of subsystems.
A key ingredient of this method is the PT-MPO which encodes the influence of the environment and reduces it to the most physically relevant sector of the state space.
Compressing the environment influence before tackling the full many-body problem greatly reduces the effective dimension of the computation and makes the simulation of a large class of many-body open quantum systems numerically feasible.

Using this approach we studied two examples.  Our first example showed how the FDT can reveal when a non-equilibrium steady state has in fact reached a thermalized state.  We also showed how it can reveal features of the non-thermalized state that occurs in the presence of a temperature gradient.   Our second example showed the ability of the method we present to model more complex situations with baths coupled to every site.

We note that the PT-MPO approach can be applied to various other many-body tensor network methods~\cite{Verstraete2004a, Vidal2007a, Orus2007, Murg2010, Zaletel2015} and that it can be modified to compute full-counting statistics of heat transfer~\cite{Popovic2021}.
It thus enables the development of a versatile set of numerical tools to study dynamics, correlations, and thermodynamic properties of many-body open quantum systems.

\begin{acknowledgments}
G.E.F. and D.K. acknowledge support from EPSRC (EP/L015110/1).
B.W.L. and J.K. acknowledge support from EPSRC (EP/T014032/1).
\end{acknowledgments}

\appendix

\section{Process tensors and TEBD}
\label{app:PT-TEBD}

In this section we present  details of our numerical method, that combines the process tensor approach to open quantum systems with time evolving block decimation (TEBD).
This method allows us to compute multi-time correlations of 1D many-body quantum systems in the presence of strongly coupled and structured environments.
We assume a total Hamiltonian of the form
\begin{equation}
	\label{eq:total-hamiltonian}
	\op{H} = \sum_{n=1}^{N} \left( \op{H}^S_n + \op{H}^E_n \right) + \sum_{n=1}^{N-1} \op{K}_{n,n+1} \mathrm{.}
\end{equation}
It consists of on-site system Hamiltonians \mbox{$\op{H}^S_n$}, on-site system-environment interaction parts \mbox{$\op{H}^E_n$}, and nearest neighbor coupling terms \mbox{$\op{K}_{n,n+1}$} for each of the \mbox{$N$} sites.
Formally, we require each of these operators to be in the set of bounded linear operators
$\mathcal{B}(\mathcal{H})$  on the appropriate Hilbert space $\mathcal{H}$.
If $\mathcal{H}^S_n$ and $\mathcal{H}^E_n$ denote the system and environment Hilbert spaces of the $n^\mathrm{th}$ site, then \mbox{$\op{H}^S_n \in \mathcal{B}(\mathcal{H}^S_n)$},  \mbox{$\op{H}^E_n \in \mathcal{B}(\mathcal{H}^S_n \otimes \mathcal{H}^E_n)$}, and  \mbox{$\op{K}_{n,n+1} \in \mathcal{B}(\mathcal{H}^S_n \otimes \mathcal{H}^S_{n+1})$}.

In principle the on-site system Hamiltonians can be completely absorbed in the definition of the on-site system-environment interactions.
However, we will see below that it is often useful to separate the pure system part from the interaction part as much as possible.
In addition to the total Hamiltonian in Eq.~\eqref{eq:total-hamiltonian}, we will also allow for on-site time-local dissipative processes described by a local master equation of GKSL form.

In the following we derive the construction of the TEBD tensor network augmented with the process tensor approach and present a suitable contraction algorithm.
Finally, we will show how to extract the intermediate time chain dynamics as well as multi-site multi-time correlations.

\subsection{Tensor network construction}
\begin{figure*}
	\includegraphics[width=0.99\textwidth]{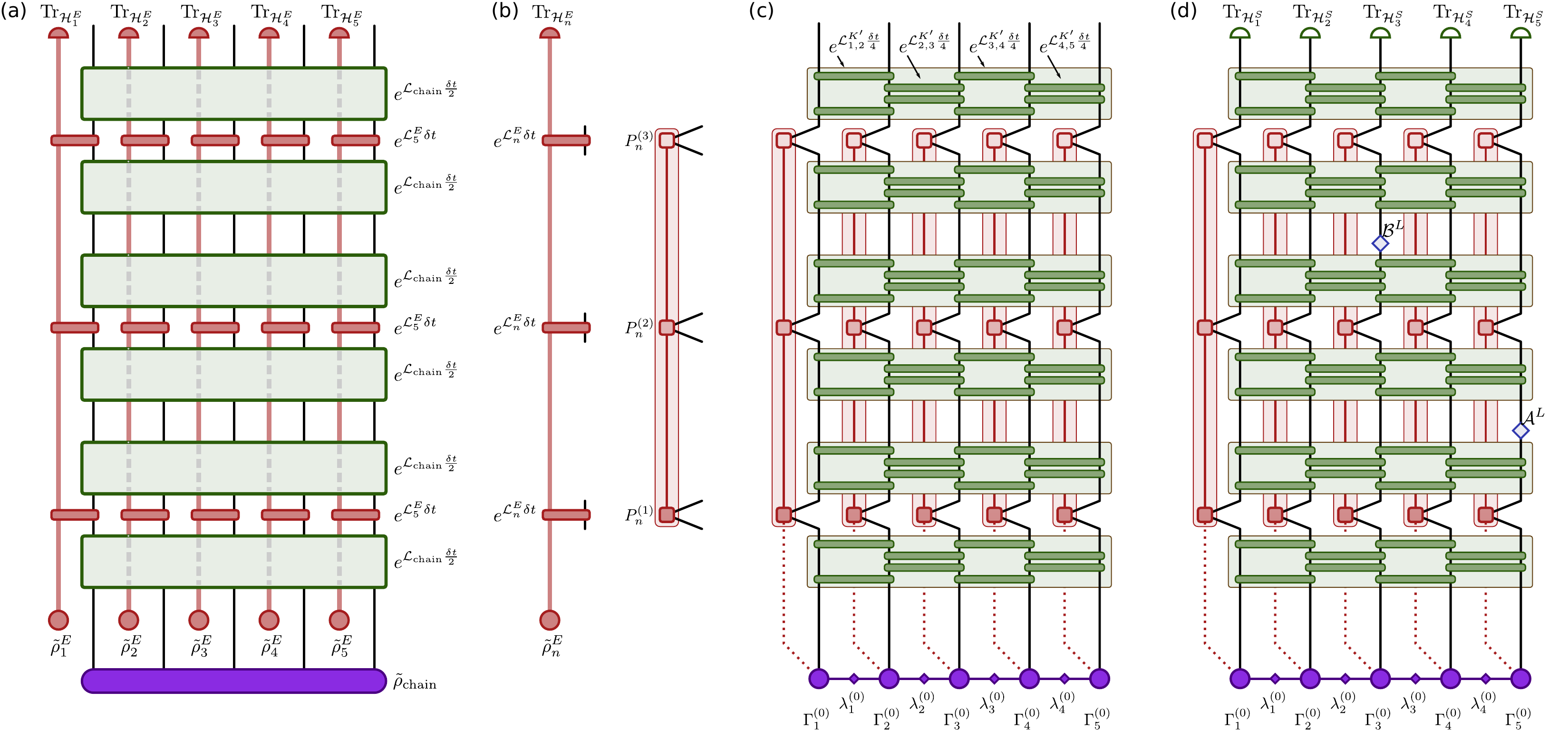}
	\caption{\label{fig:tensor-network-construction}%
		Tensor networks combining process tensors and TEBD.
		(a)~Tensor network for three time steps using a second-order Suzuki-Trotter splitting between a 5-site chain and its environments.
		(b)~A process tensor (left hand side) and the corresponding PT-MPO (right hand side).
		(c)~Full tensor network for a 5-site chain using a second order Suzuki-Trotter splitting in both environment and inter-site coupling.
		(d)~Full tensor network to compute the two time correlation \mbox{$\expect{\op{B}(2\Dt), \op{A}(1\Dt)}$}.}
\end{figure*}%
The entire following calculation is carried out in Liouville space, i.e. we consider super-operators that act on the space of vectorized density matrices.
As a start we consider the formal solution of the von Neumann equation for the total density operator at time $t$
\begin{equation}
	\rho(t) = \e{\LL t} \rho(0)
\end{equation}
with the total Liouvillian $\LL = -i [ \op{H},\cdot ]$.
We can separate the total Liouvillian into a chain and environment part
\begin{equation}
	\LL = \LL_\mathrm{chain} + \sum_{n=1}^{N} \LL^E_n \mathrm{,}
\end{equation}
where
\begin{equation}
	\LL_\mathrm{chain} = \sum_{n=1}^N \LL^S_n + \sum_{n=1}^{N-1} \LL^K_{n,n+1} \mathrm{,}
\end{equation}
with each Liouvillian corresponding to a part of the total Hamiltonian.
As mentioned above, the system Liouvillians may additionally include dissipative terms, i.e.
\begin{equation}
	\LL_n^S\,\cdot = -i[\op{H}^S_n,\cdot] + \sum_k \left( \op{L}_{n,k}^\dagger \cdot \op{L}_{n,k} -\frac{1}{2}\{\op{L}_{n,k}^\dagger \op{L}_{n,k}, \cdot \} \right)\mathrm{,}
\end{equation}
with GKSL operators $\op{L}_{n,k} \in \mathcal{B}(\mathcal{H}_n^S)$.

As a first approximation, we divide the total propagation into $M$ short time steps $\Dt$ and perform a second-order Suzuki-Trotter splitting~\cite{Suzuki1992} between the chain and the environment terms
\begin{align}
	\e{\LL t}
	&= \spare{ \e{\LL \Dt} }^M \\
	&\simeq \spare{ \e{\LL_\mathrm{chain}\frac{\Dt}{2}}\e{\pare{\sum_n^{N} \LL^E_n \Dt}} \e{\LL_\mathrm{chain}\frac{\Dt}{2}} }^M \\
	&= \spare{ \e{\LL_\mathrm{chain}\frac{\Dt}{2}} \pare{\prod_{n=1}^{N} \e{\LL^E_n \Dt}}  \e{\LL_\mathrm{chain}\frac{\Dt}{2}} }^M \mathrm{,} \label{eq:propagator-env}
\end{align}
where the last equality follows from the fact that the $\LL^E_n$ act on disjoint spaces.
For ease of presentation we now apply this propagator to a total initial state that is separable between the chain and each environment, i.e. $\rho(0) = \tilde{\rho}_\mathrm{chain} \bigotimes_n \tilde{\rho}_n^E$.
We comment below how this can be extended to initially correlated states.
Let us now consider the reduced chain state, which we obtain by performing the partial traces over all environments, i.e. $ \mathrm{Tr}_E := \mathrm{Tr}_{\{\bigotimes_n \mathcal{H}_n^E \}}$.
Assuming an initial separable state and the approximated propagator from Eq.~\eqref{eq:propagator-env}, Fig.~\ref{fig:tensor-network-construction}a expresses $\rho_\mathrm{chain}(t) := \mathrm{Tr}_E \e{\LL t} \rho(0)$ as a tensor network for three time steps.

So far, the tensor network in Fig.~\ref{fig:tensor-network-construction}a is unsuitable for carrying out a numerical computation.
The tensors representing the interaction with the environment (red tensors) explicitly involve the environment Hilbert spaces, and the chain propagators $\e{\LL_\mathrm{chain}\frac{\Dt}{2}}$ (green tensors) are still assumed to be exact, with a total dimension of $\mathrm{dim}(\mathcal{H}_n^S)^{4N}$.
This is impractical for any generic environment and any chain of significant length.
In the following we present how to construct the tensor network in Fig.~\ref{fig:tensor-network-construction}c instead, which is then suitable for an efficient numerical computation.

First, we consider the part of the tensor network that consists of the environment initial state $\tilde{\rho}_n^E$, the interaction propagators $\e{\LL^E_n \Dt}$, and the final environment trace $\mathrm{Tr}_{\mathcal{H}_n^E}$, which we draw again on the left hand side of~Fig.~\ref{fig:tensor-network-construction}b.
Together, these tensors constitute a multi-linear map, which is---per definition---the process tensor of the system-environment interaction Hamiltonian $\op{H}_n^E$ for the initial state $\tilde{\rho}_n^E$~\cite{Pollock2018}.
In principle each process tensor is of dimension $\mathrm{dim}(\mathcal{H}_n^S)^{4M}$, but in many cases can be efficiently represented by a matrix product operator (MPO) with a truncated bond dimension.
This truncated bond dimension depends on the specific Hamiltonian $\op{H}^E_n$ and can be understood as a measure of quantum non-Markovianity for the environment interaction at hand~\cite{Pollock2018a}.
We have briefly described a few of the currently available methods for the computation of process tensors in MPO form (PT-MPO) for various environments~\cite{Jorgensen2019, Strathearn2019, Fux2021, Cygorek2021, Ye2021, White2021, OQuPy2022} in the main text.
The right hand side of Fig.~\ref{fig:tensor-network-construction}b shows such a PT-MPO for three time steps.
In Fig.~\ref{fig:tensor-network-construction}c we have replaced the process tensors with the PT-MPOs obtained by one of the methods mentioned above.
Given that the computation of such a PT-MPO is often numerically involved, it can be beneficial to absorb all pure on-site system terms into $\LL_n^S$ and reuse the PT-MPO for any identical occurrences of the $\LL_n^E$ environment interactions.

Next, we consider the chain propagators $\e{\LL_\mathrm{chain}\frac{\Dt}{2}}$.
To decompose these large tensors into smaller tensors, we perform another second-order Suzuki-Trotter splitting (this time among the chain sites), making use of the fact that the chain Hamiltonian only contains on-site and nearest neighbor terms.
Higher order expansions and long range couplings are also possible \cite{Suzuki1992, Zaletel2015}.
This standard procedure leads to a TEBD tensor network for the chain evolution in Liouville space~\cite{Daley2004, Zwolak2004}.
For this, we first absorb the on-site system Hamiltonians $\op{H}^S_n$ into the nearest neighbor terms $\op{K}_{n,n+1}$ by defining
\begin{equation}
	\op{K}^\prime_{n,n+1} := \op{K}_{n,n+1} +
	\begin{cases}
		\op{H}^S_1 + \frac{1}{2}\op{H}^S_2 & \mathrm{for}\:n=1 \\
		\frac{1}{2}\op{H}^S_{N-1} + \op{H}^S_N & \mathrm{for}\:n=N-1 \\
		\frac{1}{2}\op{H}^S_n + \frac{1}{2}\op{H}^S_{n+1} & \mathrm{otherwise,}
	\end{cases}
\end{equation}
such that $\LL_\mathrm{chain} = \sum_{n=1}^{N-1} \LL^{K^\prime}_{n,n+1}$.
A first-order Suzuki-Trotter splitting, for example, then yields
\begin{align}
	\e{\LL_\mathrm{chain}\frac{\Dt}{2}}
	&= \E{\sum_n \LL_{n,n+1}^{K^\prime} \frac{\Dt}{2}} \\
	&= \E{\sum_{n\,\mathrm{odd}} \LL_{n,n+1}^{K^\prime} \frac{\Dt}{2} + \sum_{n\,\mathrm{even}} \LL_{n,n+1}^{K^\prime} \frac{\Dt}{2} } \\
	&\simeq \prod_{n\,\mathrm{odd}} \e{\LL_{n,n+1}^{K^\prime} \frac{\Dt}{2}} \prod_{n\,\mathrm{even}} \e{\LL_{n,n+1}^{K^\prime} \frac{\Dt}{2}} \mathrm{.} \label{eq:first-order-trotter}
\end{align}
In Fig.~\ref{fig:tensor-network-construction}c we replace the half time step chain propagators $\e{\LL_\mathrm{chain}\frac{\Dt}{2}}$ with a second-order Suzuki-Trotter splitting, which is of a similar form as the first-order splitting presented in equation Eq.~\eqref{eq:first-order-trotter} and consists of two body gates of the form $\e{\LL^{K^\prime}_{n,n+1} \frac{\Dt}{4}}$.

Finally, we insert the initial chain state as a matrix product state (MPS) in Vidal form~\cite{Vidal2004}.
Although the pure chain propagation works completely analogously to TEBD, each of the $\Gamma$ tensors of the MPS needs to have an extra leg which corresponds to the entanglement of the chain site with its environment.
We will call this MPS the \emph{augmented MPS}.
For initially uncorrelated chain-environment states the MPS has initially no such legs.
In such cases, we nonetheless include dummy legs of dimension 1.
These are indicated with dotted lines in Fig~\ref{fig:tensor-network-construction}c.
We do this such that the contraction algorithm for the first time step is of the same form as for all later steps, as well as to include the case of an initially correlated state~\cite{Pollock2018}, for which the dimension of the dotted lines \mbox{is $> 1$}.

\subsection{Contraction algorithm}
\begin{figure*}
	\includegraphics[width=0.99\textwidth]{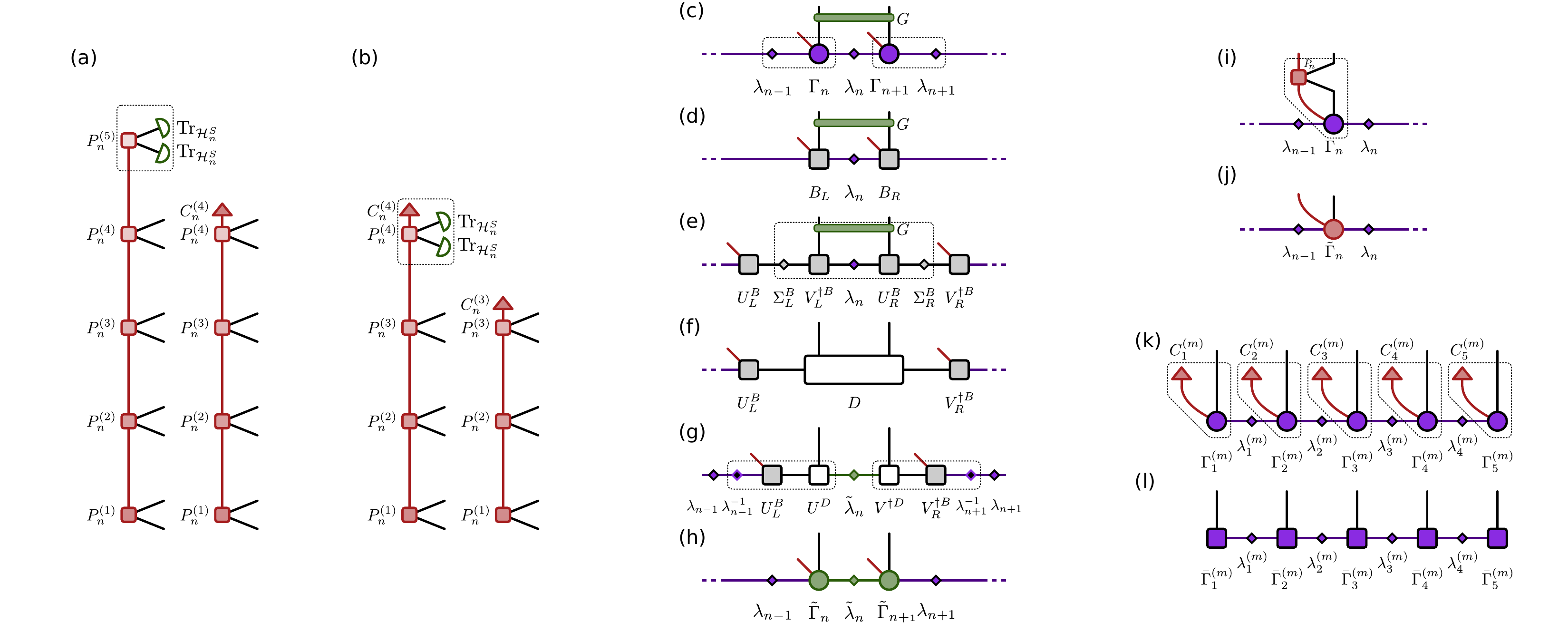}
	\caption{\label{fig:tensor-network-contraction}%
		Contraction algorithms for the PT-MPO and augmented MPS.
		(a)~Construction of the $4^\mathrm{th}$ cap tensor $C^{(4)}_n$ of the PT-MPO at site $n$.
		(b)~Construction of the $3^\mathrm{rd}$ cap tensor $C^{(3)}_n$.
		(c-h)~Contraction and decomposition sequence for the application of a two-site gate $G$ on the augmented MPS.
		(i-j)~Contraction of the augmented MPS with a PT-MPO tensor.
		(k-l)~Contraction of the augmented MPS with the cap tensors, yielding a canonical MPS in Liouville space.}
\end{figure*}%
Figure~\ref{fig:tensor-network-construction}c shows the full tensor network for three time steps of a 5-site chain.
To contract such a network we propose to absorb the tensors into the augmented MPS line by line.
This involves two different types of contraction sequences which we describe in the following.

The first type is a contraction of the augmented MPS with the chain propagators, which consist of two-site nearest neighbor gates.
We suggest a sequence of operations in \mbox{Figs.~\ref{fig:tensor-network-contraction}(c-h)}.
Compared to the canonical TEBD, this sequence includes some additional operations for the augmented legs with the aim of minimizing the size of the intermediate tensors involved.
Figures~\ref{fig:tensor-network-contraction}(c-h) show the proposed operations for applying a two body gate \mbox{$G = \e{\LL^{K^\prime}_{n,n+1} \frac{\Dt}{4}}$} to an augmented MPS:
\begin{itemize}
	\item[(c-d)] Contraction: \\ $B_L := \lambda_{n-1} \Gamma_{n}$ and $B_R := \Gamma_{n+1} \lambda_{n+1}$
	\item[(d-e)] Truncated singular value decomposition: \\ $U^B_L \Sigma^B_L V^{\dagger B}_L :\simeq B_L$ and $U^B_R \Sigma^B_R V^{\dagger B}_R :\simeq B_R$
	\item[(e-f)] Contraction: \\ $D := \Sigma^B_L \, V^{\dagger B}_L \, \lambda_{n} \, G \, U^B_R \, \Sigma^B_R$
	\item[(f-g)] Truncated singular value decomposition: \\ $U^D \Sigma^D V^{\dagger D} :\simeq D$ and $\tilde{\lambda}_{n} := \Sigma^D$
	\item[(g)] Insert identities: \\ $\lambda_{n-1} \lambda_{n-1}^{-1} = \id$ and $\lambda_{n+1}^{-1} \lambda_{n+1} = \id$
	\item[(g-h)] Contraction: \\ $\tilde{\Gamma}_{n} := \lambda_{n-1}^{-1} U^B_L U^D$ and $\tilde{\Gamma}_{n+1} := V^{\dagger D} V^{\dagger B}_R \lambda_{n+1}^{-1}$
\end{itemize}
We use a relative singular value truncation threshold~$\epsilon$, which we typically set to be of the order $10^{-6}$.
Let \mbox{$\tilde{U} \tilde{\Sigma} \tilde{V}^{\dagger} = X$} be an exact singular value decomposition of the matrix \mbox{$X \in \mathbb{C}^{a \times b}$}.
We then choose the truncated bond dimension $\chi$ to be as small as possible, while maintaining \mbox{$\norm{\tilde{\Sigma} - \Sigma}_2 < \epsilon \max[\tilde{\Sigma}]$}.
Here, \mbox{$U \in \mathbb{C}^{a \times \chi}$}, \mbox{$\Sigma \in \mathbb{C}^{\chi \times \chi}$}, and \mbox{$V^{\dagger} \in \mathbb{C}^{\chi \times b}$} denote the truncated matrices and $\norm{\tilde{\Sigma} - \Sigma}_2$ is the 2-norm of the discarded singular values.

The other type of operation that occurs when absorbing the tensor network line by line is the contraction of the augmented MPS with the subsequent parts of the PT-MPOs.
\mbox{Figures~\ref{fig:tensor-network-contraction}(i-j)} show the contraction of an augmented MPS site ($\Gamma_n$) with a single tensor of a PT-MPO ($P_n$).
This contraction only updates the $\Gamma$ tensors of the augmented MPS, where the bond legs of the PT-MPOs become the new augmented legs of the augmented MPS.

We point out that the contraction sequences described above only act locally on a short part of the augmented MPS for each step.
This contraction scheme is therefore well suited for parallel computing.

\subsection{Intermediate chain evolution}
As presented thus far, this method would only yield a reduced chain state at the final time step.
We can, however, extract the reduced density matrix of the chain for every intermediate time step by temporarily removing the correlations of the augmented MPS with the environment.
This can be done using the so called \emph{containment} property of process tensors, which allows the generation of process tensors for a smaller set of time slots by tracing over all later time slots \cite{Pollock2018}.
For this we construct the tensors $C_n^{(m)}$ (which we call \emph{cap} tensors) as shown in \mbox{Figs.~\ref{fig:tensor-network-contraction}(a-b)}.
Applying these cap tensors to the augmented MPS at time \mbox{step $m$} removes the augmented leg and yields a canonical MPS that represents the vectorized reduced density matrix of the chain at that time (see Figs.~\ref{fig:tensor-network-contraction}(k-l)).

\subsection{Multi-site multi-time correlations}
Finally, we explain how to compute multi-site multi-time correlations.
As an example, we could be interested in the correlation $\expect{\op{B}(2\Dt), \op{A}(1\Dt)}$, with $\op{A}$ and $\op{B}$ acting on the $5^\mathrm{th}$ and $3^\mathrm{rd}$ spin of a \mbox{5-site} chain respectively.
More generally, we consider all correlations $C$ of the form
\begin{equation}
	C = \expect{ \prod_{p=1}^P \op{C}_p(m_p \Dt) } \mathrm{,}
\end{equation}
with $P$ time-ordered operators acting on possibly different chain sites $\op{C}_p \in \mathcal{B}(\mathcal{H}_{n_p}^S)$ at times $m_p \Dt$.
This can be written as
\begin{align}
	C
	&= \mathrm{Tr}\spare{ \prod_{p=1}^P \pare{ \mathcal{C}^L_p \e{\LL (m_p - m_{p-1})\Dt} } \rho(0) } \\
	&= \mathrm{Tr}\spare{ \prod_{p=1}^P \pare{ \mathcal{C}^L_p [\e{\LL \Dt}]^{(m_p - m_{p-1})} } \rho(0) } \mathrm{,} \label{eq:two-time-corr-decomposed}
\end{align}
with \mbox{$m_0 := 0$} and the left acting super-operators \mbox{$\mathcal{C}^L_p := \op{C}_p\,\cdot$}.
To represent Eq.~\eqref{eq:two-time-corr-decomposed} as a tensor network, we replace the full propagators $\e{\LL \Dt}$ with the same construction as above.
This leads to the same tensor network as in Fig.~\ref{fig:tensor-network-construction}c, but with additionally inserted super-operators and with additional traces over the chain sites at the top of the network.
We exemplify this in Fig.~\ref{fig:tensor-network-construction}d for the above example of \mbox{$\expect{\op{B}(2\Dt), \op{A}(1\Dt)}$}.
Finally, we mention that for out-of-time ordered correlations, the operators need to be inserted as right acting super-operators \mbox{$\mathcal{C}^R_p := \cdot\,\op{C}_p$} instead.

\section{Details of calculations for XYZ spin chain with thermal leads}
\label{app:XYZ-details}

In this section we present further details and results of the spin chain simulations discussed in the main text.
We comment on the process tensor computation and explain how we use the PT-MPO approach to TEBD introduced in the previous section to compute the fluctuation and dissipation spectra.

\subsection{Process tensor computation}
The method introduced above requires pre-computed PT-MPOs to capture the interactions with the environments.
We consider bosonic baths with an Ohmic spectral density, coupling strength of $\alpha=0.32$, and a cutoff frequency $\omega_c=4.0$ (see Eq.~\eqref{eq:bosonic-bath} and the text following).
A suitable method for the computation of the corresponding PT-MPO is the a process tensor adoption of the time evolving matrix product operator (PT-TEMPO) method~\cite{Jorgensen2019, Strathearn2019, Fux2021, OQuPy2022}.
Such a computation has three convergence parameters: the time step $\Dt$, the maximal number of memory steps $\Delta K_\mathrm{max}$, and the relative singular value truncation threshold $\epsilon_\mathrm{TEMPO}$.
The product $\Dt \, \Delta K_\mathrm{max}$ is the maximal correlation time of the environment that is included in the computation.
The environment correlation function
\begin{equation}
	C(\tau) = \int_0^{\infty} J(\omega) \left[ \cos(\omega \tau) \coth\left( \frac{\omega}{2 T}\right) - i \sin(\omega \tau) \right] \mathrm{d}\omega
\end{equation}
drops at time $t=8.0$ below $10^{-3}$ of its maximum value.
Consistent with this, we find the choice of $\Dt=0.2$, $\Delta K_\mathrm{max} = 40$, and $\epsilon_\mathrm{TEMPO} = 10^{-6}$ to be adequate.
We comment on checking the convergence of the simulations in more detail further below.
We carried out the computation of the PT-MPOs using the open source package OQuPy~\cite{OQuPy2022}.
It took approximately 4~minutes to calculate a process tensor with 1600 time steps on a single core of an Intel i7 (8th Gen) processor.
The resulting process tensors have a maximal bond dimension of 37 and 44 for the temperatures $T_\mathrm{hot}=1.6$ and $T_\mathrm{cold}=0.8$ respectively.

\subsection{Multi-time correlations and fluctuation and dissipation spectra}
To extract dissipation and fluctuation spectra we first evolve the spin chain from the initial state at time $t_i=0.0$ to the approximate steady state at time $t_{ss}$. During the TEBD propagation we used a relative singular value truncation of $\epsilon_\mathrm{TEBD} = 10^{-6}$.
We found that for all scenarios considered, $t_{ss}=192.0$ (960 steps) is long enough to reach an approximate steady state.
For the results presented in this Paper we have chosen the initial state of each spin to be $\tilde{\rho}_n^S \propto \E{-\frac{\epsilon_n \op{s}^z}{2 T_n}}$ where $T_n = T_\mathrm{hot}$ and \mbox{$T_n = T_\mathrm{hot} + \frac{n-1}{8} (T_\mathrm{cold}-T_\mathrm{hot})$} for the 5-site and the 9-site chains respectively.
As expected, we found the same steady state and the same two-time correlations when starting from other random initial product states.

Then, to compute two-time correlations such as \mbox{$\expect{\op{B}(t_{ss}+\tau) \op{A}(t_{ss})}$} with respect to some single site operators $\op{A}$ and $\op{B}$, we apply the left acting super-operator $\mathcal{A}^L=\op{A} \cdot$ to the steady state and compute the expectation value of $\op{B}$ for all later times up to the final time $t_f=320.0$.
We can thus compute the two time correlations $\expect{\op{\sigma}^z_n(t_{ss}+\tau) \op{\sigma}^z_n(t_{ss})}$ of spin $n$ for all $\tau$ up to $\tau_\mathrm{max}:=t_f-t_{ss}$ with a single propagation starting from the steady state.
It is important to point out that the expression ``steady state'' refers to the state of the whole (chain and environments) and not just the reduced chain state.
Because the two-time auto-correlations of the chain depend on the steady state correlations of the chain with the environment, it is vital to continue the propagation from the full augmented MPS, incorporating entanglement with the environment,  at time $t_{ss}$.

With this, we obtain $\expect{\op{\sigma}^z_n(t_{ss}+\tau) \op{\sigma}^z_n(t_{ss})}$ for all \mbox{$\tau \in (0, \tau_\mathrm{max})$}, which we identify with $\expect{\op{\sigma}^z_n(\tau) \op{\sigma}^z_n(0)}_{ss}$ at the steady state.
Using
\begin{align}
	\expect{\op{\sigma}^z_n(\tau) \op{\sigma}^z_n(0)}_{ss}
	&= \expect{\op{\sigma}^z_n(0) \op{\sigma}^z_n(-\tau)}_{ss} \nonumber\\
	&= \expect{\op{\sigma}^z_n(0) \op{\sigma}^z_n(\tau)}^{*}_{ss}
\end{align}
we can construct commutators and anti-commutators for $\tau\in(-\tau_\mathrm{max}, \tau_\mathrm{max})$ and employ a fast Fourier transformation on this interval to compute the fluctuation and dissipation spectra.

In order to check the convergence of the simulations with respect to the computation parameters, we study the finite differences of our results with respect to altered parameters.
We performed simulations substituting \mbox{$\Dt = 0.2 \rightarrow 0.15$}, \mbox{$\Delta K_\mathrm{max}=40 \rightarrow 30$}, \mbox{$\epsilon_\mathrm{TEMPO} = 10^{-6} \rightarrow 10^{-5}$} for the process tensor computation; and \mbox{$\epsilon_\mathrm{TEBD} = 10^{-6} \rightarrow 10^{-5}$}, \mbox{$t_{ss}=192.0 \rightarrow 160.0$}, and \mbox{$\tau_\mathrm{max}=128.0 \rightarrow 160.0$} for the augmented TEBD evolution.
We found that the resulting differences are dominated by the variation of $\epsilon_\mathrm{TEBD}$ and we thus use these differences as an estimate for the numerical error.
We plot this error estimate in Figs.~\ref{fig:spin-chain-N9-z-with-chain}b~and~\ref{fig:spin-chain-N9-z-with-chain}d by broadening the lines appropriately.
For all other plots (see Figs.~\ref{fig:spin-chain-N5-z3-with-chain},\,\ref{fig:spin-chain-N9-x},~and~\ref{fig:spin-chain-N9-y}) the estimated error is smaller or similar to the thickness of the lines.

We carried out all computations employing code that we have made public as part of the open source python package OQuPy~\cite{OQuPy2022}.
The total propagation for the single bath 5-site chain took 68~minutes on a single core of an Intel i7 (8th Gen) machine.
For the 9-site chain the propagation from the initial state to the steady state took 8~hours~6~minutes on four cores of an Intel \mbox{Xeon E5-2695} machine.
The propagation after the application of the first $\op{\sigma}^z_n$ took between 6~hours~37~minutes and 8~hours~39~minutes, depending on the site $n$ to which it was applied.

\subsection{MPS bond dimension during two-time calculations}
\begin{figure}[tb]
	\includegraphics[width=0.49\textwidth]{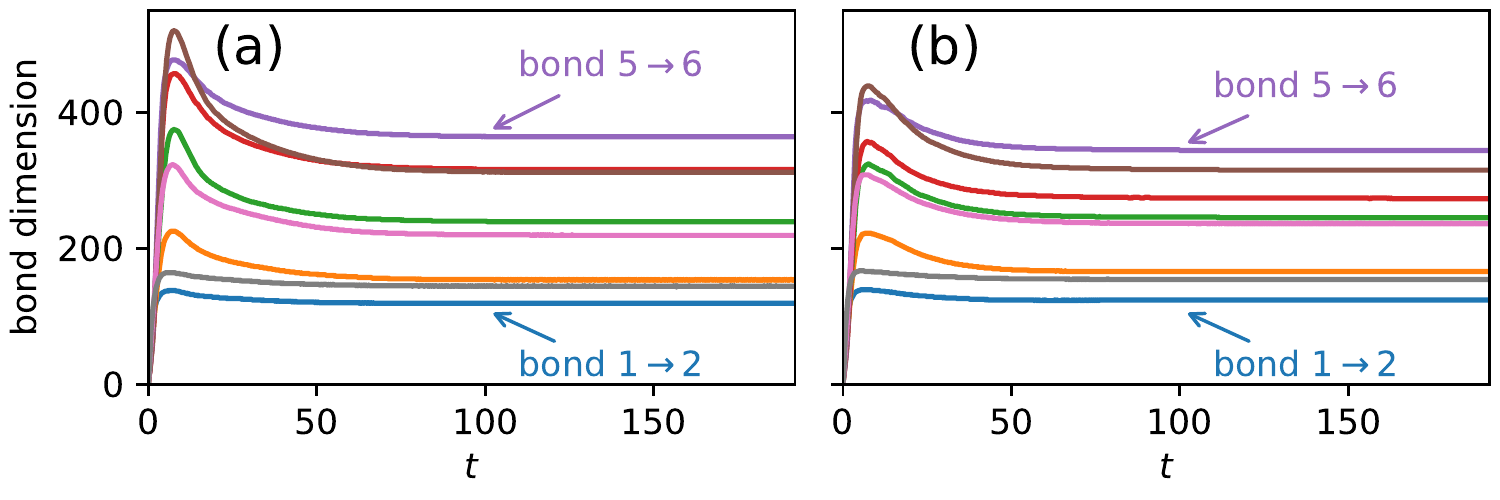}
	\caption{\label{fig:bond-dims-N9-ss}%
		The bond dimensions of the augmented MPS during the propagation of the 9-site spin chain placed between two baths from the initial to the approximate steady state, for the clean (a) and disordered case (b).}
\end{figure}%
\begin{figure}[tb]
	\includegraphics[width=0.49\textwidth]{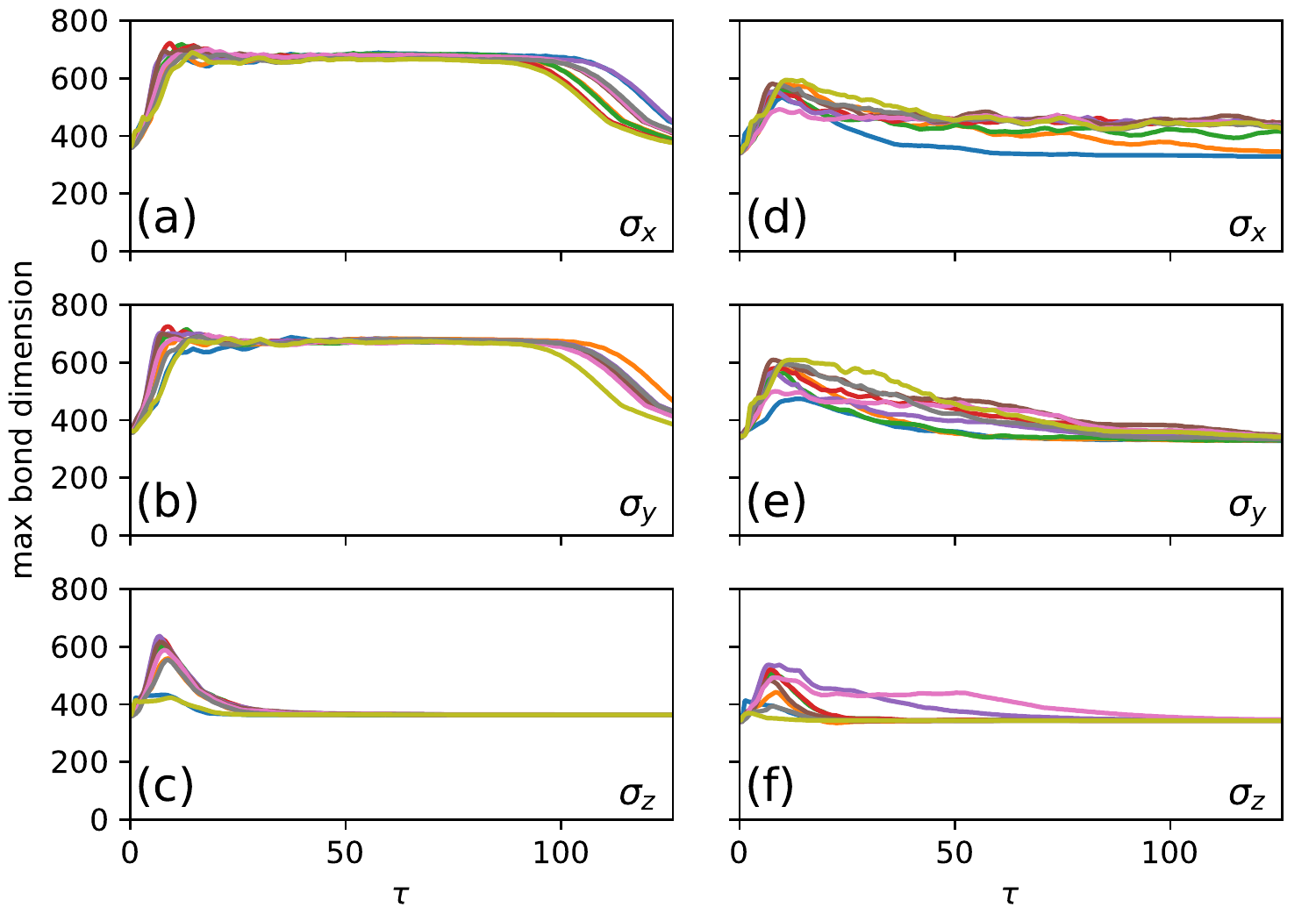}
	\caption{\label{fig:bond-dims-N9-ttc}%
		The maximal bond dimension of the 9-site spin chain augmented MPS as a function of time after the application of $\op{\sigma}^x_n$, $\op{\sigma}^y_n$, and $\op{\sigma}^z_n$ for each site $n$ and the clean (a-c) and disordered case (d-f).}
\end{figure}%
Figure~\ref{fig:bond-dims-N9-ss} shows the bond dimensions of the augmented MPS during the propagation of the spin chain from the initial to the approximate steady state.
As expected, the bond dimension is larger towards the middle of the chain.
It appears that the transient dynamics of the chain passes through a state with a significantly higher entanglement entropy among the sites compared to the steady state, signaled by the bond dimension peak at about $t=8.0$.
Panel~(c) of Fig.~\ref{fig:bond-dims-N9-ttc} shows the maximal bond dimension of the augmented MPS as a function of time after the application of $\op{\sigma}^z_n$ for each site $n$ of the clean spin chain.
It shows a similar behavior to the bond dimension during the initial propagation and peaks at approximately $\tau=7.0$.
In Figs.~\ref{fig:bond-dims-N9-ttc}a~and~\ref{fig:bond-dims-N9-ttc}b we observe a plateau at its highest value before it drops towards the end of the simulation.

\section{Details of calculation for 21-site XY spin chain}
\label{app:xy-details}

We choose \mbox{$\Dt = 0.2$}, \mbox{$\Delta K_\mathrm{max}=40$}, \mbox{$\epsilon_\mathrm{TEMPO} = 10^{-6}$} for the process tensor computation and find converged dynamics for PT-MPOs with a small bond dimensions of $\xi=26$ and $\xi=37$ for $\alpha=0.16$ and $\alpha=0.32$, respectively.

The computation of the chain dynamics through the augmented TEBD tensor network strongly depends on the anisotropy $\eta$.
For a given relative singular value truncation threshold $\epsilon_\mathrm{TEBD}$ the maximal necessary bond dimension $\chi$ increases with larger anisotropy.
Considering the entire dynamics for coupling strength $\alpha=0.32$ and choosing $\epsilon_\mathrm{TEBD}=10^{-6}$ we find $\chi=204$ and $\chi=382$ for $\eta=0.0$ and $\eta=0.04$, respectively.
While for the isotropic case $\chi$ settles to a constant value after the initial quench, we find that $\chi$ slowly grows (approximately linearly) over time for the anisotropic case.
This is consistent with the intuition that the additional excitations in the system due to the anisotropy lead to a growth of the subspace in which the most relevant spin dynamics takes place.
Further simulations suggest that even for the anisotropic case the bond dimension $\chi$ is, however, independent of the chain length, which means that the computation scales approximately linearly with system size.
The most challenging computation numerically was the chain dynamics for $\alpha=0.32$ and $\eta=0.04$, for which the entire computation took 46 hours on a single computing node using 32 Intel \mbox{Xeon E5-2695} CPUs.

\end{document}